\documentstyle[11pt]{article}

\newcommand{\be}{\begin{equation}}
\newcommand{\ee}{\end{equation}}
\newcommand{\bea}{\begin{eqnarray}}
\newcommand{\eea}{\end{eqnarray}}

\newcommand{\ks}{{\mbox k \!\!\! /}}
\newcommand{\ps}{{\mbox p \!\!\! /}}
\newcommand{\qs}{{\mbox q \!\!\! /}}

\newcommand{\NPB}[3]{Nucl.\ Phys.\ {\bf B{#1}} (19{#2}) {#3}}
\newcommand{\PRD}[3]{Phys.\ Rev.\ {\bf D{#1}} (19{#2}) {#3}}
\newcommand{\PLB}[3]{Phys.\ Lett.\ {\bf B{#1}} (19{#2}) {#3}}

\newsavebox{\rgluonarrow}
\savebox{\rgluonarrow}(0,0)[l]{
\thicklines
\put(11.5,2){\line(-1,0){4}}
\put(11.5,2){\line(0,-1){4}}
\thinlines
}

\newsavebox{\lgluonarrow}
\savebox{\lgluonarrow}(0,0)[r]{
\thicklines
\put(-11.5,-2){\line(1,0){4}}
\put(-11.5,-2){\line(0,1){4}}
\thinlines
}

\newsavebox{\ugluonarrow}
\savebox{\ugluonarrow}(0,0)[b]{
\thicklines
\put(-2,11.5){\line(1,0){4}}
\put(-2,11.5){\line(0,-1){4}}
\thinlines
}

\newsavebox{\urgluonarrow}
\savebox{\urgluonarrow}(0,0)[b]{
\thicklines
\put( 2,6.5){\line(-1,0){4}}
\put( 2,6.5){\line(0,-1){4}}
\thinlines
}

\newsavebox{\hblob}
\savebox{\hblob}(0,0)[c]{
\put(0,0){\oval(40,20)}
\put(-10,-10){\line(1,1){20}}
\put(-5,-10){\line(1,1){19.1}}
\put( 0,-10){\line(1,1){17.1}}
\put( 5,-10){\line(1,1){14.1}}
\put(10,-10){\line(1,1){10}}
\put( 5, 10){\line(-1,-1){19.1}}
\put( 0, 10){\line(-1,-1){17.1}}
\put(-5, 10){\line(-1,-1){14.1}}
\put(-10, 10){\line(-1,-1){10}}
}

\newsavebox{\roundblob}
\savebox{\roundblob}(0,0)[c]{
\put(  0,0){\circle{40}}
\put(-19.1,5.9){\line(1,1){13.2}}
\put(-10,10){\line(1,1){10}}
\put(-10,10){\line(-1,-1){10}}
\put(-7.5,7.5){\line(1,1){12}}
\put(-7.5,7.5){\line(-1,-1){12}}
\put(-5, 5){\line(1,1){13.2}}
\put(-5, 5){\line(-1,-1){13.2}}
\put(-2.5,2.5){\line(-1,-1){13.9}}
\put(-2.5,2.5){\line(1,1){13.9}}
\put( 0,0){\line(1,1){14.1}}
\put( 0,0){\line(-1,-1){14.1}}
\put(2.5,-2.5){\line(-1,-1){13.9}}
\put(2.5,-2.5){\line(1,1){13.9}}
\put( 5, -5){\line(1,1){13.2}}
\put( 5, -5){\line(-1,-1){13.2}}
\put(7.5,-7.5){\line(1,1){12}}
\put(7.5,-7.5){\line(-1,-1){12}}
\put(10,-10){\line(1,1){10}}
\put(10,-10){\line(-1,-1){10}}
\put(5.9,-19.1){\line(1,1){13.2}}
}

\newsavebox{\rightloop}
\savebox{\rightloop}(0,0)[c]{
\put(  0.00, 20.00){\circle*{0.2}}
\put(  0.40, 19.92){\circle*{0.2}}
\put(  0.79, 19.70){\circle*{0.2}}
\put(  1.16, 19.34){\circle*{0.2}}
\put(  1.51, 18.87){\circle*{0.2}}
\put(  1.84, 18.31){\circle*{0.2}}
\put(  2.13, 17.70){\circle*{0.2}}
\put(  2.40, 17.06){\circle*{0.2}}
\put(  2.65, 16.43){\circle*{0.2}}
\put(  2.88, 15.85){\circle*{0.2}}
\put(  3.11, 15.34){\circle*{0.2}}
\put(  3.34, 14.94){\circle*{0.2}}
\put(  3.59, 14.65){\circle*{0.2}}
\put(  3.86, 14.50){\circle*{0.2}}
\put(  4.16, 14.47){\circle*{0.2}}
\put(  4.51, 14.58){\circle*{0.2}}
\put(  4.90, 14.79){\circle*{0.2}}
\put(  5.34, 15.11){\circle*{0.2}}
\put(  5.83, 15.48){\circle*{0.2}}
\put(  6.35, 15.90){\circle*{0.2}}
\put(  6.90, 16.32){\circle*{0.2}}
\put(  7.46, 16.71){\circle*{0.2}}
\put(  8.03, 17.05){\circle*{0.2}}
\put(  8.57, 17.30){\circle*{0.2}}
\put(  9.08, 17.44){\circle*{0.2}}
\put(  9.54, 17.46){\circle*{0.2}}
\put(  9.94, 17.35){\circle*{0.2}}
\put( 10.26, 17.11){\circle*{0.2}}
\put( 10.50, 16.75){\circle*{0.2}}
\put( 10.66, 16.27){\circle*{0.2}}
\put( 10.74, 15.70){\circle*{0.2}}
\put( 10.75, 15.06){\circle*{0.2}}
\put( 10.71, 14.38){\circle*{0.2}}
\put( 10.63, 13.69){\circle*{0.2}}
\put( 10.53, 13.02){\circle*{0.2}}
\put( 10.44, 12.39){\circle*{0.2}}
\put( 10.37, 11.83){\circle*{0.2}}
\put( 10.36, 11.34){\circle*{0.2}}
\put( 10.41, 10.96){\circle*{0.2}}
\put( 10.55, 10.67){\circle*{0.2}}
\put( 10.79, 10.48){\circle*{0.2}}
\put( 11.12, 10.38){\circle*{0.2}}
\put( 11.56, 10.36){\circle*{0.2}}
\put( 12.08, 10.40){\circle*{0.2}}
\put( 12.67, 10.48){\circle*{0.2}}
\put( 13.33, 10.58){\circle*{0.2}}
\put( 14.01, 10.67){\circle*{0.2}}
\put( 14.70, 10.74){\circle*{0.2}}
\put( 15.36, 10.75){\circle*{0.2}}
\put( 15.97, 10.71){\circle*{0.2}}
\put( 16.50, 10.60){\circle*{0.2}}
\put( 16.93, 10.40){\circle*{0.2}}
\put( 17.24, 10.12){\circle*{0.2}}
\put( 17.42,  9.76){\circle*{0.2}}
\put( 17.47,  9.34){\circle*{0.2}}
\put( 17.39,  8.85){\circle*{0.2}}
\put( 17.20,  8.32){\circle*{0.2}}
\put( 16.90,  7.77){\circle*{0.2}}
\put( 16.54,  7.20){\circle*{0.2}}
\put( 16.13,  6.64){\circle*{0.2}}
\put( 15.71,  6.11){\circle*{0.2}}
\put( 15.30,  5.60){\circle*{0.2}}
\put( 14.95,  5.13){\circle*{0.2}}
\put( 14.68,  4.72){\circle*{0.2}}
\put( 14.51,  4.34){\circle*{0.2}}
\put( 14.47,  4.02){\circle*{0.2}}
\put( 14.55,  3.73){\circle*{0.2}}
\put( 14.77,  3.47){\circle*{0.2}}
\put( 15.11,  3.23){\circle*{0.2}}
\put( 15.57,  3.01){\circle*{0.2}}
\put( 16.11,  2.78){\circle*{0.2}}
\put( 16.72,  2.54){\circle*{0.2}}
\put( 17.35,  2.28){\circle*{0.2}}
\put( 17.99,  2.00){\circle*{0.2}}
\put( 18.58,  1.69){\circle*{0.2}}
\put( 19.10,  1.35){\circle*{0.2}}
\put( 19.52,  0.99){\circle*{0.2}}
\put( 19.82,  0.61){\circle*{0.2}}
\put( 19.98,  0.22){\circle*{0.2}}
\put( 19.98, -0.18){\circle*{0.2}}
\put( 19.84, -0.58){\circle*{0.2}}
\put( 19.55, -0.96){\circle*{0.2}}
\put( 19.14, -1.33){\circle*{0.2}}
\put( 18.62, -1.67){\circle*{0.2}}
\put( 18.04, -1.98){\circle*{0.2}}
\put( 17.40, -2.26){\circle*{0.2}}
\put( 16.77, -2.52){\circle*{0.2}}
\put( 16.16, -2.76){\circle*{0.2}}
\put( 15.61, -2.99){\circle*{0.2}}
\put( 15.14, -3.22){\circle*{0.2}}
\put( 14.79, -3.45){\circle*{0.2}}
\put( 14.56, -3.71){\circle*{0.2}}
\put( 14.47, -3.99){\circle*{0.2}}
\put( 14.51, -4.32){\circle*{0.2}}
\put( 14.66, -4.68){\circle*{0.2}}
\put( 14.93, -5.10){\circle*{0.2}}
\put( 15.27, -5.56){\circle*{0.2}}
\put( 15.67, -6.06){\circle*{0.2}}
\put( 16.09, -6.60){\circle*{0.2}}
\put( 16.51, -7.16){\circle*{0.2}}
\put( 16.88, -7.72){\circle*{0.2}}
\put( 17.18, -8.28){\circle*{0.2}}
\put( 17.38, -8.81){\circle*{0.2}}
\put( 17.47, -9.30){\circle*{0.2}}
\put( 17.43, -9.73){\circle*{0.2}}
\put( 17.26,-10.09){\circle*{0.2}}
\put( 16.96,-10.38){\circle*{0.2}}
\put( 16.54,-10.58){\circle*{0.2}}
\put( 16.02,-10.71){\circle*{0.2}}
\put( 15.41,-10.75){\circle*{0.2}}
\put( 14.75,-10.74){\circle*{0.2}}
\put( 14.07,-10.68){\circle*{0.2}}
\put( 13.38,-10.58){\circle*{0.2}}
\put( 12.73,-10.49){\circle*{0.2}}
\put( 12.12,-10.40){\circle*{0.2}}
\put( 11.59,-10.36){\circle*{0.2}}
\put( 11.15,-10.37){\circle*{0.2}}
\put( 10.81,-10.47){\circle*{0.2}}
\put( 10.57,-10.65){\circle*{0.2}}
\put( 10.42,-10.93){\circle*{0.2}}
\put( 10.36,-11.31){\circle*{0.2}}
\put( 10.37,-11.79){\circle*{0.2}}
\put( 10.43,-12.34){\circle*{0.2}}
\put( 10.52,-12.97){\circle*{0.2}}
\put( 10.62,-13.64){\circle*{0.2}}
\put( 10.70,-14.33){\circle*{0.2}}
\put( 10.75,-15.01){\circle*{0.2}}
\put( 10.74,-15.65){\circle*{0.2}}
\put( 10.67,-16.23){\circle*{0.2}}
\put( 10.51,-16.71){\circle*{0.2}}
\put( 10.28,-17.09){\circle*{0.2}}
\put(  9.96,-17.34){\circle*{0.2}}
\put(  9.57,-17.46){\circle*{0.2}}
\put(  9.12,-17.45){\circle*{0.2}}
\put(  8.61,-17.32){\circle*{0.2}}
\put(  8.07,-17.07){\circle*{0.2}}
\put(  7.51,-16.74){\circle*{0.2}}
\put(  6.94,-16.35){\circle*{0.2}}
\put(  6.39,-15.93){\circle*{0.2}}
\put(  5.87,-15.52){\circle*{0.2}}
\put(  5.38,-15.13){\circle*{0.2}}
\put(  4.94,-14.82){\circle*{0.2}}
\put(  4.54,-14.59){\circle*{0.2}}
\put(  4.19,-14.48){\circle*{0.2}}
\put(  3.88,-14.49){\circle*{0.2}}
\put(  3.61,-14.64){\circle*{0.2}}
\put(  3.36,-14.91){\circle*{0.2}}
\put(  3.13,-15.31){\circle*{0.2}}
\put(  2.90,-15.81){\circle*{0.2}}
\put(  2.67,-16.38){\circle*{0.2}}
\put(  2.42,-17.01){\circle*{0.2}}
\put(  2.16,-17.65){\circle*{0.2}}
\put(  1.86,-18.27){\circle*{0.2}}
\put(  1.54,-18.83){\circle*{0.2}}
\put(  1.19,-19.31){\circle*{0.2}}
\put(  0.82,-19.68){\circle*{0.2}}
\put(  0.43,-19.91){\circle*{0.2}}
}

\newsavebox{\leftloop}
\savebox{\leftloop}(0,0)[c]{
\put(  0.03,-20.00){\circle*{0.2}}
\put( -0.37,-19.94){\circle*{0.2}}
\put( -0.76,-19.72){\circle*{0.2}}
\put( -1.13,-19.38){\circle*{0.2}}
\put( -1.49,-18.91){\circle*{0.2}}
\put( -1.81,-18.36){\circle*{0.2}}
\put( -2.11,-17.75){\circle*{0.2}}
\put( -2.38,-17.11){\circle*{0.2}}
\put( -2.63,-16.48){\circle*{0.2}}
\put( -2.87,-15.89){\circle*{0.2}}
\put( -3.09,-15.38){\circle*{0.2}}
\put( -3.32,-14.97){\circle*{0.2}}
\put( -3.57,-14.67){\circle*{0.2}}
\put( -3.83,-14.50){\circle*{0.2}}
\put( -4.14,-14.47){\circle*{0.2}}
\put( -4.48,-14.56){\circle*{0.2}}
\put( -4.87,-14.77){\circle*{0.2}}
\put( -5.31,-15.08){\circle*{0.2}}
\put( -5.79,-15.45){\circle*{0.2}}
\put( -6.31,-15.87){\circle*{0.2}}
\put( -6.86,-16.29){\circle*{0.2}}
\put( -7.42,-16.68){\circle*{0.2}}
\put( -7.98,-17.03){\circle*{0.2}}
\put( -8.53,-17.28){\circle*{0.2}}
\put( -9.04,-17.44){\circle*{0.2}}
\put( -9.51,-17.47){\circle*{0.2}}
\put( -9.91,-17.37){\circle*{0.2}}
\put(-10.24,-17.14){\circle*{0.2}}
\put(-10.48,-16.78){\circle*{0.2}}
\put(-10.65,-16.31){\circle*{0.2}}
\put(-10.74,-15.75){\circle*{0.2}}
\put(-10.75,-15.11){\circle*{0.2}}
\put(-10.71,-14.44){\circle*{0.2}}
\put(-10.64,-13.75){\circle*{0.2}}
\put(-10.54,-13.07){\circle*{0.2}}
\put(-10.44,-12.44){\circle*{0.2}}
\put(-10.38,-11.87){\circle*{0.2}}
\put(-10.36,-11.38){\circle*{0.2}}
\put(-10.41,-10.98){\circle*{0.2}}
\put(-10.54,-10.69){\circle*{0.2}}
\put(-10.77,-10.49){\circle*{0.2}}
\put(-11.09,-10.38){\circle*{0.2}}
\put(-11.52,-10.36){\circle*{0.2}}
\put(-12.03,-10.39){\circle*{0.2}}
\put(-12.63,-10.47){\circle*{0.2}}
\put(-13.27,-10.57){\circle*{0.2}}
\put(-13.96,-10.66){\circle*{0.2}}
\put(-14.64,-10.73){\circle*{0.2}}
\put(-15.31,-10.76){\circle*{0.2}}
\put(-15.92,-10.72){\circle*{0.2}}
\put(-16.46,-10.61){\circle*{0.2}}
\put(-16.90,-10.42){\circle*{0.2}}
\put(-17.22,-10.14){\circle*{0.2}}
\put(-17.41, -9.79){\circle*{0.2}}
\put(-17.47, -9.37){\circle*{0.2}}
\put(-17.40, -8.89){\circle*{0.2}}
\put(-17.22, -8.37){\circle*{0.2}}
\put(-16.93, -7.81){\circle*{0.2}}
\put(-16.57, -7.25){\circle*{0.2}}
\put(-16.16, -6.69){\circle*{0.2}}
\put(-15.74, -6.15){\circle*{0.2}}
\put(-15.33, -5.64){\circle*{0.2}}
\put(-14.98, -5.17){\circle*{0.2}}
\put(-14.70, -4.75){\circle*{0.2}}
\put(-14.52, -4.37){\circle*{0.2}}
\put(-14.47, -4.04){\circle*{0.2}}
\put(-14.54, -3.75){\circle*{0.2}}
\put(-14.75, -3.49){\circle*{0.2}}
\put(-15.08, -3.25){\circle*{0.2}}
\put(-15.53, -3.02){\circle*{0.2}}
\put(-16.06, -2.80){\circle*{0.2}}
\put(-16.67, -2.56){\circle*{0.2}}
\put(-17.30, -2.30){\circle*{0.2}}
\put(-17.94, -2.02){\circle*{0.2}}
\put(-18.54, -1.72){\circle*{0.2}}
\put(-19.06, -1.38){\circle*{0.2}}
\put(-19.50, -1.02){\circle*{0.2}}
\put(-19.80, -0.64){\circle*{0.2}}
\put(-19.97, -0.25){\circle*{0.2}}
\put(-19.99,  0.15){\circle*{0.2}}
\put(-19.86,  0.55){\circle*{0.2}}
\put(-19.58,  0.93){\circle*{0.2}}
\put(-19.18,  1.30){\circle*{0.2}}
\put(-18.67,  1.64){\circle*{0.2}}
\put(-18.08,  1.95){\circle*{0.2}}
\put(-17.46,  2.24){\circle*{0.2}}
\put(-16.82,  2.50){\circle*{0.2}}
\put(-16.20,  2.74){\circle*{0.2}}
\put(-15.65,  2.97){\circle*{0.2}}
\put(-15.18,  3.20){\circle*{0.2}}
\put(-14.82,  3.43){\circle*{0.2}}
\put(-14.58,  3.69){\circle*{0.2}}
\put(-14.47,  3.97){\circle*{0.2}}
\put(-14.50,  4.29){\circle*{0.2}}
\put(-14.65,  4.65){\circle*{0.2}}
\put(-14.90,  5.06){\circle*{0.2}}
\put(-15.24,  5.52){\circle*{0.2}}
\put(-15.64,  6.02){\circle*{0.2}}
\put(-16.06,  6.56){\circle*{0.2}}
\put(-16.47,  7.11){\circle*{0.2}}
\put(-16.85,  7.68){\circle*{0.2}}
\put(-17.16,  8.24){\circle*{0.2}}
\put(-17.37,  8.77){\circle*{0.2}}
\put(-17.47,  9.26){\circle*{0.2}}
\put(-17.44,  9.70){\circle*{0.2}}
\put(-17.28, 10.07){\circle*{0.2}}
\put(-16.99, 10.36){\circle*{0.2}}
\put(-16.58, 10.57){\circle*{0.2}}
\put(-16.06, 10.70){\circle*{0.2}}
\put(-15.46, 10.75){\circle*{0.2}}
\put(-14.81, 10.74){\circle*{0.2}}
\put(-14.12, 10.68){\circle*{0.2}}
\put(-13.44, 10.59){\circle*{0.2}}
\put(-12.78, 10.49){\circle*{0.2}}
\put(-12.17, 10.41){\circle*{0.2}}
\put(-11.63, 10.36){\circle*{0.2}}
\put(-11.18, 10.37){\circle*{0.2}}
\put(-10.83, 10.46){\circle*{0.2}}
\put(-10.58, 10.63){\circle*{0.2}}
\put(-10.43, 10.90){\circle*{0.2}}
\put(-10.36, 11.28){\circle*{0.2}}
\put(-10.37, 11.74){\circle*{0.2}}
\put(-10.42, 12.30){\circle*{0.2}}
\put(-10.51, 12.92){\circle*{0.2}}
\put(-10.61, 13.59){\circle*{0.2}}
\put(-10.70, 14.27){\circle*{0.2}}
\put(-10.75, 14.96){\circle*{0.2}}
\put(-10.75, 15.60){\circle*{0.2}}
\put(-10.68, 16.18){\circle*{0.2}}
\put(-10.53, 16.68){\circle*{0.2}}
\put(-10.30, 17.06){\circle*{0.2}}
\put( -9.99, 17.32){\circle*{0.2}}
\put( -9.61, 17.46){\circle*{0.2}}
\put( -9.16, 17.46){\circle*{0.2}}
\put( -8.66, 17.33){\circle*{0.2}}
\put( -8.11, 17.10){\circle*{0.2}}
\put( -7.55, 16.77){\circle*{0.2}}
\put( -6.99, 16.39){\circle*{0.2}}
\put( -6.44, 15.97){\circle*{0.2}}
\put( -5.91, 15.55){\circle*{0.2}}
\put( -5.42, 15.16){\circle*{0.2}}
\put( -4.97, 14.84){\circle*{0.2}}
\put( -4.57, 14.60){\circle*{0.2}}
\put( -4.21, 14.48){\circle*{0.2}}
\put( -3.90, 14.48){\circle*{0.2}}
\put( -3.63, 14.62){\circle*{0.2}}
\put( -3.38, 14.89){\circle*{0.2}}
\put( -3.15, 15.27){\circle*{0.2}}
\put( -2.92, 15.76){\circle*{0.2}}
\put( -2.69, 16.34){\circle*{0.2}}
\put( -2.44, 16.96){\circle*{0.2}}
\put( -2.18, 17.60){\circle*{0.2}}
\put( -1.89, 18.22){\circle*{0.2}}
\put( -1.57, 18.79){\circle*{0.2}}
\put( -1.22, 19.28){\circle*{0.2}}
\put( -0.85, 19.65){\circle*{0.2}}
\put( -0.46, 19.90){\circle*{0.2}}
\put( -0.06, 20.00){\circle*{0.2}}
}

\newsavebox{\gluon}
\savebox{\gluon}(0,0)[c]{
\put(  0.00,  0.00){\circle*{0.2}}
\put(  0.16,  0.25){\circle*{0.2}}
\put(  0.32,  0.50){\circle*{0.2}}
\put(  0.48,  0.74){\circle*{0.2}}
\put(  0.64,  0.97){\circle*{0.2}}
\put(  0.80,  1.20){\circle*{0.2}}
\put(  0.96,  1.41){\circle*{0.2}}
\put(  1.11,  1.61){\circle*{0.2}}
\put(  1.27,  1.79){\circle*{0.2}}
\put(  1.43,  1.96){\circle*{0.2}}
\put(  1.59,  2.10){\circle*{0.2}}
\put(  1.75,  2.23){\circle*{0.2}}
\put(  1.91,  2.33){\circle*{0.2}}
\put(  2.07,  2.41){\circle*{0.2}}
\put(  2.23,  2.46){\circle*{0.2}}
\put(  2.39,  2.49){\circle*{0.2}}
\put(  2.55,  2.50){\circle*{0.2}}
\put(  2.71,  2.48){\circle*{0.2}}
\put(  2.87,  2.43){\circle*{0.2}}
\put(  3.03,  2.37){\circle*{0.2}}
\put(  3.18,  2.27){\circle*{0.2}}
\put(  3.34,  2.16){\circle*{0.2}}
\put(  3.50,  2.02){\circle*{0.2}}
\put(  3.66,  1.86){\circle*{0.2}}
\put(  3.82,  1.69){\circle*{0.2}}
\put(  3.98,  1.50){\circle*{0.2}}
\put(  4.14,  1.29){\circle*{0.2}}
\put(  4.30,  1.07){\circle*{0.2}}
\put(  4.46,  0.84){\circle*{0.2}}
\put(  4.62,  0.60){\circle*{0.2}}
\put(  4.78,  0.35){\circle*{0.2}}
\put(  4.94,  0.10){\circle*{0.2}}
\put(  5.10, -0.15){\circle*{0.2}}
\put(  5.25, -0.39){\circle*{0.2}}
\put(  5.41, -0.64){\circle*{0.2}}
\put(  5.57, -0.88){\circle*{0.2}}
\put(  5.73, -1.11){\circle*{0.2}}
\put(  5.89, -1.32){\circle*{0.2}}
\put(  6.05, -1.53){\circle*{0.2}}
\put(  6.21, -1.72){\circle*{0.2}}
\put(  6.37, -1.89){\circle*{0.2}}
\put(  6.53, -2.05){\circle*{0.2}}
\put(  6.69, -2.18){\circle*{0.2}}
\put(  6.85, -2.29){\circle*{0.2}}
\put(  7.01, -2.38){\circle*{0.2}}
\put(  7.17, -2.44){\circle*{0.2}}
\put(  7.32, -2.48){\circle*{0.2}}
\put(  7.48, -2.50){\circle*{0.2}}
\put(  7.64, -2.49){\circle*{0.2}}
\put(  7.80, -2.46){\circle*{0.2}}
\put(  7.96, -2.40){\circle*{0.2}}
\put(  8.12, -2.31){\circle*{0.2}}
\put(  8.28, -2.21){\circle*{0.2}}
\put(  8.44, -2.08){\circle*{0.2}}
\put(  8.60, -1.93){\circle*{0.2}}
\put(  8.76, -1.76){\circle*{0.2}}
\put(  8.92, -1.58){\circle*{0.2}}
\put(  9.08, -1.38){\circle*{0.2}}
\put(  9.24, -1.16){\circle*{0.2}}
\put(  9.39, -0.93){\circle*{0.2}}
\put(  9.55, -0.70){\circle*{0.2}}
\put(  9.71, -0.46){\circle*{0.2}}
\put(  9.87, -0.21){\circle*{0.2}}
\put( 10.03,  0.04){\circle*{0.2}}
\put( 10.19,  0.29){\circle*{0.2}}
\put( 10.35,  0.54){\circle*{0.2}}
\put( 10.51,  0.78){\circle*{0.2}}
\put( 10.67,  1.01){\circle*{0.2}}
\put( 10.83,  1.24){\circle*{0.2}}
\put( 10.99,  1.45){\circle*{0.2}}
\put( 11.15,  1.64){\circle*{0.2}}
\put( 11.31,  1.82){\circle*{0.2}}
\put( 11.46,  1.98){\circle*{0.2}}
\put( 11.62,  2.13){\circle*{0.2}}
\put( 11.78,  2.25){\circle*{0.2}}
\put( 11.94,  2.34){\circle*{0.2}}
\put( 12.10,  2.42){\circle*{0.2}}
\put( 12.26,  2.47){\circle*{0.2}}
\put( 12.42,  2.50){\circle*{0.2}}
\put( 12.58,  2.50){\circle*{0.2}}
\put( 12.74,  2.47){\circle*{0.2}}
\put( 12.90,  2.42){\circle*{0.2}}
\put( 13.06,  2.35){\circle*{0.2}}
\put( 13.22,  2.26){\circle*{0.2}}
\put( 13.38,  2.14){\circle*{0.2}}
\put( 13.54,  2.00){\circle*{0.2}}
\put( 13.69,  1.84){\circle*{0.2}}
\put( 13.85,  1.66){\circle*{0.2}}
\put( 14.01,  1.46){\circle*{0.2}}
\put( 14.17,  1.25){\circle*{0.2}}
\put( 14.33,  1.03){\circle*{0.2}}
\put( 14.49,  0.80){\circle*{0.2}}
\put( 14.65,  0.56){\circle*{0.2}}
\put( 14.81,  0.31){\circle*{0.2}}
\put( 14.97,  0.06){\circle*{0.2}}
\put( 15.13, -0.19){\circle*{0.2}}
\put( 15.29, -0.44){\circle*{0.2}}
\put( 15.45, -0.68){\circle*{0.2}}
\put( 15.61, -0.92){\circle*{0.2}}
\put( 15.76, -1.14){\circle*{0.2}}
\put( 15.92, -1.36){\circle*{0.2}}
\put( 16.08, -1.56){\circle*{0.2}}
\put( 16.24, -1.75){\circle*{0.2}}
\put( 16.40, -1.92){\circle*{0.2}}
\put( 16.56, -2.07){\circle*{0.2}}
\put( 16.72, -2.20){\circle*{0.2}}
\put( 16.88, -2.31){\circle*{0.2}}
\put( 17.04, -2.39){\circle*{0.2}}
\put( 17.20, -2.45){\circle*{0.2}}
\put( 17.36, -2.49){\circle*{0.2}}
\put( 17.52, -2.50){\circle*{0.2}}
\put( 17.68, -2.49){\circle*{0.2}}
\put( 17.83, -2.45){\circle*{0.2}}
\put( 17.99, -2.39){\circle*{0.2}}
\put( 18.15, -2.30){\circle*{0.2}}
\put( 18.31, -2.19){\circle*{0.2}}
\put( 18.47, -2.06){\circle*{0.2}}
\put( 18.63, -1.90){\circle*{0.2}}
\put( 18.79, -1.73){\circle*{0.2}}
\put( 18.95, -1.55){\circle*{0.2}}
\put( 19.11, -1.34){\circle*{0.2}}
\put( 19.27, -1.12){\circle*{0.2}}
\put( 19.43, -0.90){\circle*{0.2}}
\put( 19.59, -0.66){\circle*{0.2}}
\put( 19.75, -0.41){\circle*{0.2}}
\put( 19.90, -0.17){\circle*{0.2}}
}

\newsavebox{\vgluon}
\savebox{\vgluon}(0,0)[b]{
\put(  0.00,  0.00){\circle*{0.2}}
\put( -0.20,  0.13){\circle*{0.2}}
\put( -0.40,  0.25){\circle*{0.2}}
\put( -0.59,  0.38){\circle*{0.2}}
\put( -0.79,  0.51){\circle*{0.2}}
\put( -0.97,  0.64){\circle*{0.2}}
\put( -1.15,  0.76){\circle*{0.2}}
\put( -1.33,  0.89){\circle*{0.2}}
\put( -1.49,  1.02){\circle*{0.2}}
\put( -1.65,  1.15){\circle*{0.2}}
\put( -1.79,  1.27){\circle*{0.2}}
\put( -1.93,  1.40){\circle*{0.2}}
\put( -2.05,  1.53){\circle*{0.2}}
\put( -2.16,  1.66){\circle*{0.2}}
\put( -2.25,  1.78){\circle*{0.2}}
\put( -2.33,  1.91){\circle*{0.2}}
\put( -2.40,  2.04){\circle*{0.2}}
\put( -2.44,  2.17){\circle*{0.2}}
\put( -2.48,  2.29){\circle*{0.2}}
\put( -2.50,  2.42){\circle*{0.2}}
\put( -2.50,  2.55){\circle*{0.2}}
\put( -2.49,  2.68){\circle*{0.2}}
\put( -2.46,  2.80){\circle*{0.2}}
\put( -2.41,  2.93){\circle*{0.2}}
\put( -2.35,  3.06){\circle*{0.2}}
\put( -2.27,  3.18){\circle*{0.2}}
\put( -2.18,  3.31){\circle*{0.2}}
\put( -2.08,  3.44){\circle*{0.2}}
\put( -1.96,  3.57){\circle*{0.2}}
\put( -1.83,  3.69){\circle*{0.2}}
\put( -1.69,  3.82){\circle*{0.2}}
\put( -1.54,  3.95){\circle*{0.2}}
\put( -1.37,  4.08){\circle*{0.2}}
\put( -1.20,  4.20){\circle*{0.2}}
\put( -1.02,  4.33){\circle*{0.2}}
\put( -0.84,  4.46){\circle*{0.2}}
\put( -0.65,  4.59){\circle*{0.2}}
\put( -0.45,  4.71){\circle*{0.2}}
\put( -0.25,  4.84){\circle*{0.2}}
\put( -0.05,  4.97){\circle*{0.2}}
\put(  0.15,  5.10){\circle*{0.2}}
\put(  0.34,  5.22){\circle*{0.2}}
\put(  0.54,  5.35){\circle*{0.2}}
\put(  0.73,  5.48){\circle*{0.2}}
\put(  0.92,  5.61){\circle*{0.2}}
\put(  1.11,  5.73){\circle*{0.2}}
\put(  1.28,  5.86){\circle*{0.2}}
\put(  1.45,  5.99){\circle*{0.2}}
\put(  1.61,  6.11){\circle*{0.2}}
\put(  1.76,  6.24){\circle*{0.2}}
\put(  1.89,  6.37){\circle*{0.2}}
\put(  2.02,  6.50){\circle*{0.2}}
\put(  2.13,  6.62){\circle*{0.2}}
\put(  2.23,  6.75){\circle*{0.2}}
\put(  2.31,  6.88){\circle*{0.2}}
\put(  2.38,  7.01){\circle*{0.2}}
\put(  2.43,  7.13){\circle*{0.2}}
\put(  2.47,  7.26){\circle*{0.2}}
\put(  2.49,  7.39){\circle*{0.2}}
\put(  2.50,  7.52){\circle*{0.2}}
\put(  2.49,  7.64){\circle*{0.2}}
\put(  2.46,  7.77){\circle*{0.2}}
\put(  2.42,  7.90){\circle*{0.2}}
\put(  2.37,  8.03){\circle*{0.2}}
\put(  2.30,  8.15){\circle*{0.2}}
\put(  2.21,  8.28){\circle*{0.2}}
\put(  2.11,  8.41){\circle*{0.2}}
\put(  1.99,  8.54){\circle*{0.2}}
\put(  1.87,  8.66){\circle*{0.2}}
\put(  1.73,  8.79){\circle*{0.2}}
\put(  1.58,  8.92){\circle*{0.2}}
\put(  1.42,  9.04){\circle*{0.2}}
\put(  1.25,  9.17){\circle*{0.2}}
\put(  1.07,  9.30){\circle*{0.2}}
\put(  0.89,  9.43){\circle*{0.2}}
\put(  0.70,  9.55){\circle*{0.2}}
\put(  0.50,  9.68){\circle*{0.2}}
\put(  0.31,  9.81){\circle*{0.2}}
\put(  0.11,  9.94){\circle*{0.2}}
\put( -0.09, 10.06){\circle*{0.2}}
\put( -0.29, 10.19){\circle*{0.2}}
\put( -0.49, 10.32){\circle*{0.2}}
\put( -0.68, 10.45){\circle*{0.2}}
\put( -0.87, 10.57){\circle*{0.2}}
\put( -1.06, 10.70){\circle*{0.2}}
\put( -1.24, 10.83){\circle*{0.2}}
\put( -1.41, 10.96){\circle*{0.2}}
\put( -1.57, 11.08){\circle*{0.2}}
\put( -1.72, 11.21){\circle*{0.2}}
\put( -1.86, 11.34){\circle*{0.2}}
\put( -1.98, 11.46){\circle*{0.2}}
\put( -2.10, 11.59){\circle*{0.2}}
\put( -2.20, 11.72){\circle*{0.2}}
\put( -2.29, 11.85){\circle*{0.2}}
\put( -2.36, 11.97){\circle*{0.2}}
\put( -2.42, 12.10){\circle*{0.2}}
\put( -2.46, 12.23){\circle*{0.2}}
\put( -2.49, 12.36){\circle*{0.2}}
\put( -2.50, 12.48){\circle*{0.2}}
\put( -2.49, 12.61){\circle*{0.2}}
\put( -2.47, 12.74){\circle*{0.2}}
\put( -2.44, 12.87){\circle*{0.2}}
\put( -2.38, 12.99){\circle*{0.2}}
\put( -2.32, 13.12){\circle*{0.2}}
\put( -2.23, 13.25){\circle*{0.2}}
\put( -2.14, 13.38){\circle*{0.2}}
\put( -2.03, 13.50){\circle*{0.2}}
\put( -1.90, 13.63){\circle*{0.2}}
\put( -1.77, 13.76){\circle*{0.2}}
\put( -1.62, 13.89){\circle*{0.2}}
\put( -1.46, 14.01){\circle*{0.2}}
\put( -1.30, 14.14){\circle*{0.2}}
\put( -1.12, 14.27){\circle*{0.2}}
\put( -0.94, 14.39){\circle*{0.2}}
\put( -0.75, 14.52){\circle*{0.2}}
\put( -0.56, 14.65){\circle*{0.2}}
\put( -0.36, 14.78){\circle*{0.2}}
\put( -0.16, 14.90){\circle*{0.2}}
\put(  0.04, 15.03){\circle*{0.2}}
\put(  0.24, 15.16){\circle*{0.2}}
\put(  0.44, 15.29){\circle*{0.2}}
\put(  0.63, 15.41){\circle*{0.2}}
\put(  0.82, 15.54){\circle*{0.2}}
\put(  1.01, 15.67){\circle*{0.2}}
\put(  1.19, 15.80){\circle*{0.2}}
\put(  1.36, 15.92){\circle*{0.2}}
\put(  1.52, 16.05){\circle*{0.2}}
\put(  1.68, 16.18){\circle*{0.2}}
\put(  1.82, 16.31){\circle*{0.2}}
\put(  1.95, 16.43){\circle*{0.2}}
\put(  2.07, 16.56){\circle*{0.2}}
\put(  2.18, 16.69){\circle*{0.2}}
\put(  2.27, 16.82){\circle*{0.2}}
\put(  2.34, 16.94){\circle*{0.2}}
\put(  2.41, 17.07){\circle*{0.2}}
\put(  2.45, 17.20){\circle*{0.2}}
\put(  2.48, 17.32){\circle*{0.2}}
\put(  2.50, 17.45){\circle*{0.2}}
\put(  2.50, 17.58){\circle*{0.2}}
\put(  2.48, 17.71){\circle*{0.2}}
\put(  2.45, 17.83){\circle*{0.2}}
\put(  2.40, 17.96){\circle*{0.2}}
\put(  2.34, 18.09){\circle*{0.2}}
\put(  2.26, 18.22){\circle*{0.2}}
\put(  2.16, 18.34){\circle*{0.2}}
\put(  2.06, 18.47){\circle*{0.2}}
\put(  1.94, 18.60){\circle*{0.2}}
\put(  1.80, 18.73){\circle*{0.2}}
\put(  1.66, 18.85){\circle*{0.2}}
\put(  1.51, 18.98){\circle*{0.2}}
\put(  1.34, 19.11){\circle*{0.2}}
\put(  1.17, 19.24){\circle*{0.2}}
\put(  0.99, 19.36){\circle*{0.2}}
\put(  0.80, 19.49){\circle*{0.2}}
\put(  0.61, 19.62){\circle*{0.2}}
\put(  0.41, 19.75){\circle*{0.2}}
\put(  0.22, 19.87){\circle*{0.2}}
}

\newsavebox{\trgluon}
\savebox{\trgluon}(0,0)[bl]{
\put(  0.00,  0.00){\circle*{0.2}}
\put(  0.03,  0.26){\circle*{0.2}}
\put(  0.06,  0.53){\circle*{0.2}}
\put(  0.08,  0.79){\circle*{0.2}}
\put(  0.12,  1.04){\circle*{0.2}}
\put(  0.15,  1.29){\circle*{0.2}}
\put(  0.19,  1.53){\circle*{0.2}}
\put(  0.24,  1.76){\circle*{0.2}}
\put(  0.28,  1.98){\circle*{0.2}}
\put(  0.34,  2.19){\circle*{0.2}}
\put(  0.40,  2.39){\circle*{0.2}}
\put(  0.47,  2.57){\circle*{0.2}}
\put(  0.55,  2.74){\circle*{0.2}}
\put(  0.63,  2.89){\circle*{0.2}}
\put(  0.72,  3.02){\circle*{0.2}}
\put(  0.82,  3.14){\circle*{0.2}}
\put(  0.93,  3.24){\circle*{0.2}}
\put(  1.05,  3.32){\circle*{0.2}}
\put(  1.18,  3.38){\circle*{0.2}}
\put(  1.32,  3.43){\circle*{0.2}}
\put(  1.47,  3.45){\circle*{0.2}}
\put(  1.62,  3.46){\circle*{0.2}}
\put(  1.78,  3.45){\circle*{0.2}}
\put(  1.96,  3.43){\circle*{0.2}}
\put(  2.14,  3.39){\circle*{0.2}}
\put(  2.32,  3.33){\circle*{0.2}}
\put(  2.52,  3.27){\circle*{0.2}}
\put(  2.72,  3.18){\circle*{0.2}}
\put(  2.92,  3.09){\circle*{0.2}}
\put(  3.14,  2.98){\circle*{0.2}}
\put(  3.35,  2.87){\circle*{0.2}}
\put(  3.57,  2.75){\circle*{0.2}}
\put(  3.80,  2.62){\circle*{0.2}}
\put(  4.02,  2.49){\circle*{0.2}}
\put(  4.25,  2.35){\circle*{0.2}}
\put(  4.48,  2.22){\circle*{0.2}}
\put(  4.71,  2.08){\circle*{0.2}}
\put(  4.93,  1.94){\circle*{0.2}}
\put(  5.16,  1.81){\circle*{0.2}}
\put(  5.38,  1.68){\circle*{0.2}}
\put(  5.60,  1.56){\circle*{0.2}}
\put(  5.82,  1.45){\circle*{0.2}}
\put(  6.03,  1.35){\circle*{0.2}}
\put(  6.23,  1.26){\circle*{0.2}}
\put(  6.43,  1.18){\circle*{0.2}}
\put(  6.63,  1.11){\circle*{0.2}}
\put(  6.81,  1.06){\circle*{0.2}}
\put(  6.99,  1.02){\circle*{0.2}}
\put(  7.16,  1.00){\circle*{0.2}}
\put(  7.32,  0.99){\circle*{0.2}}
\put(  7.48,  1.01){\circle*{0.2}}
\put(  7.62,  1.04){\circle*{0.2}}
\put(  7.76,  1.09){\circle*{0.2}}
\put(  7.88,  1.15){\circle*{0.2}}
\put(  8.00,  1.24){\circle*{0.2}}
\put(  8.11,  1.34){\circle*{0.2}}
\put(  8.21,  1.46){\circle*{0.2}}
\put(  8.30,  1.60){\circle*{0.2}}
\put(  8.38,  1.75){\circle*{0.2}}
\put(  8.46,  1.92){\circle*{0.2}}
\put(  8.52,  2.10){\circle*{0.2}}
\put(  8.58,  2.30){\circle*{0.2}}
\put(  8.64,  2.51){\circle*{0.2}}
\put(  8.69,  2.74){\circle*{0.2}}
\put(  8.73,  2.97){\circle*{0.2}}
\put(  8.77,  3.21){\circle*{0.2}}
\put(  8.80,  3.46){\circle*{0.2}}
\put(  8.83,  3.72){\circle*{0.2}}
\put(  8.86,  3.98){\circle*{0.2}}
\put(  8.89,  4.24){\circle*{0.2}}
\put(  8.92,  4.51){\circle*{0.2}}
\put(  8.95,  4.77){\circle*{0.2}}
\put(  8.97,  5.03){\circle*{0.2}}
\put(  9.00,  5.29){\circle*{0.2}}
\put(  9.04,  5.55){\circle*{0.2}}
\put(  9.07,  5.79){\circle*{0.2}}
\put(  9.11,  6.03){\circle*{0.2}}
\put(  9.16,  6.26){\circle*{0.2}}
\put(  9.21,  6.48){\circle*{0.2}}
\put(  9.26,  6.69){\circle*{0.2}}
\put(  9.33,  6.88){\circle*{0.2}}
\put(  9.40,  7.06){\circle*{0.2}}
\put(  9.47,  7.22){\circle*{0.2}}
\put(  9.56,  7.37){\circle*{0.2}}
\put(  9.66,  7.50){\circle*{0.2}}
\put(  9.76,  7.61){\circle*{0.2}}
\put(  9.87,  7.71){\circle*{0.2}}
\put(  9.99,  7.79){\circle*{0.2}}
\put( 10.12,  7.85){\circle*{0.2}}
\put( 10.26,  7.89){\circle*{0.2}}
\put( 10.41,  7.91){\circle*{0.2}}
\put( 10.56,  7.92){\circle*{0.2}}
\put( 10.73,  7.91){\circle*{0.2}}
\put( 10.90,  7.88){\circle*{0.2}}
\put( 11.08,  7.84){\circle*{0.2}}
\put( 11.27,  7.78){\circle*{0.2}}
\put( 11.47,  7.71){\circle*{0.2}}
\put( 11.67,  7.62){\circle*{0.2}}
\put( 11.88,  7.53){\circle*{0.2}}
\put( 12.09,  7.42){\circle*{0.2}}
\put( 12.31,  7.30){\circle*{0.2}}
\put( 12.53,  7.18){\circle*{0.2}}
\put( 12.75,  7.05){\circle*{0.2}}
\put( 12.98,  6.92){\circle*{0.2}}
\put( 13.21,  6.78){\circle*{0.2}}
\put( 13.43,  6.65){\circle*{0.2}}
\put( 13.66,  6.51){\circle*{0.2}}
\put( 13.89,  6.38){\circle*{0.2}}
\put( 14.11,  6.24){\circle*{0.2}}
\put( 14.34,  6.12){\circle*{0.2}}
\put( 14.56,  6.00){\circle*{0.2}}
\put( 14.77,  5.89){\circle*{0.2}}
\put( 14.98,  5.79){\circle*{0.2}}
\put( 15.18,  5.70){\circle*{0.2}}
\put( 15.38,  5.62){\circle*{0.2}}
\put( 15.58,  5.56){\circle*{0.2}}
\put( 15.76,  5.51){\circle*{0.2}}
\put( 15.94,  5.47){\circle*{0.2}}
\put( 16.11,  5.45){\circle*{0.2}}
\put( 16.27,  5.45){\circle*{0.2}}
\put( 16.42,  5.47){\circle*{0.2}}
\put( 16.56,  5.50){\circle*{0.2}}
\put( 16.69,  5.55){\circle*{0.2}}
\put( 16.82,  5.62){\circle*{0.2}}
\put( 16.93,  5.71){\circle*{0.2}}
\put( 17.04,  5.82){\circle*{0.2}}
\put( 17.14,  5.94){\circle*{0.2}}
\put( 17.23,  6.08){\circle*{0.2}}
\put( 17.31,  6.24){\circle*{0.2}}
\put( 17.38,  6.41){\circle*{0.2}}
\put( 17.45,  6.60){\circle*{0.2}}
\put( 17.51,  6.80){\circle*{0.2}}
\put( 17.56,  7.01){\circle*{0.2}}
\put( 17.61,  7.24){\circle*{0.2}}
\put( 17.65,  7.47){\circle*{0.2}}
\put( 17.69,  7.71){\circle*{0.2}}
\put( 17.72,  7.97){\circle*{0.2}}
\put( 17.75,  8.22){\circle*{0.2}}
\put( 17.78,  8.48){\circle*{0.2}}
\put( 17.81,  8.75){\circle*{0.2}}
\put( 17.84,  9.01){\circle*{0.2}}
\put( 17.86,  9.28){\circle*{0.2}}
\put( 17.89,  9.54){\circle*{0.2}}
\put( 17.92,  9.80){\circle*{0.2}}
\put( 17.96, 10.05){\circle*{0.2}}
\put( 17.99, 10.30){\circle*{0.2}}
\put( 18.03, 10.53){\circle*{0.2}}
\put( 18.08, 10.76){\circle*{0.2}}
\put( 18.13, 10.98){\circle*{0.2}}
\put( 18.19, 11.18){\circle*{0.2}}
\put( 18.25, 11.37){\circle*{0.2}}
\put( 18.32, 11.55){\circle*{0.2}}
\put( 18.40, 11.71){\circle*{0.2}}
\put( 18.49, 11.85){\circle*{0.2}}
\put( 18.59, 11.98){\circle*{0.2}}
\put( 18.69, 12.09){\circle*{0.2}}
\put( 18.80, 12.18){\circle*{0.2}}
\put( 18.93, 12.26){\circle*{0.2}}
\put( 19.06, 12.31){\circle*{0.2}}
\put( 19.20, 12.35){\circle*{0.2}}
\put( 19.35, 12.37){\circle*{0.2}}
\put( 19.51, 12.38){\circle*{0.2}}
\put( 19.67, 12.36){\circle*{0.2}}
\put( 19.85, 12.33){\circle*{0.2}}
\put( 20.03, 12.28){\circle*{0.2}}
\put( 20.22, 12.22){\circle*{0.2}}
\put( 20.42, 12.15){\circle*{0.2}}
\put( 20.62, 12.06){\circle*{0.2}}
\put( 20.83, 11.96){\circle*{0.2}}
\put( 21.04, 11.86){\circle*{0.2}}
\put( 21.26, 11.74){\circle*{0.2}}
\put( 21.48, 11.61){\circle*{0.2}}
\put( 21.71, 11.49){\circle*{0.2}}
\put( 21.93, 11.35){\circle*{0.2}}
\put( 22.16, 11.21){\circle*{0.2}}
\put( 22.39, 11.08){\circle*{0.2}}
\put( 22.62, 10.94){\circle*{0.2}}
\put( 22.84, 10.81){\circle*{0.2}}
\put( 23.07, 10.68){\circle*{0.2}}
\put( 23.29, 10.55){\circle*{0.2}}
\put( 23.51, 10.43){\circle*{0.2}}
\put( 23.72, 10.33){\circle*{0.2}}
\put( 23.93, 10.23){\circle*{0.2}}
\put( 24.14, 10.14){\circle*{0.2}}
\put( 24.33, 10.06){\circle*{0.2}}
\put( 24.52, 10.00){\circle*{0.2}}
\put( 24.71,  9.96){\circle*{0.2}}
\put( 24.88,  9.92){\circle*{0.2}}
\put( 25.05,  9.91){\circle*{0.2}}
\put( 25.21,  9.91){\circle*{0.2}}
\put( 25.36,  9.93){\circle*{0.2}}
\put( 25.50,  9.97){\circle*{0.2}}
\put( 25.63, 10.02){\circle*{0.2}}
\put( 25.75, 10.09){\circle*{0.2}}
\put( 25.87, 10.19){\circle*{0.2}}
\put( 25.97, 10.29){\circle*{0.2}}
\put( 26.07, 10.42){\circle*{0.2}}
\put( 26.16, 10.56){\circle*{0.2}}
\put( 26.24, 10.72){\circle*{0.2}}
\put( 26.31, 10.90){\circle*{0.2}}
\put( 26.37, 11.09){\circle*{0.2}}
\put( 26.43, 11.29){\circle*{0.2}}
\put( 26.48, 11.51){\circle*{0.2}}
\put( 26.53, 11.73){\circle*{0.2}}
\put( 26.57, 11.97){\circle*{0.2}}
\put( 26.61, 12.22){\circle*{0.2}}
\put( 26.64, 12.47){\circle*{0.2}}
\put( 26.67, 12.73){\circle*{0.2}}
\put( 26.70, 12.99){\circle*{0.2}}
\put( 26.73, 13.25){\circle*{0.2}}
\put( 26.76, 13.52){\circle*{0.2}}
\put( 26.78, 13.78){\circle*{0.2}}
\put( 26.81, 14.04){\circle*{0.2}}
\put( 26.84, 14.30){\circle*{0.2}}
\put( 26.88, 14.55){\circle*{0.2}}
\put( 26.91, 14.80){\circle*{0.2}}
\put( 26.96, 15.03){\circle*{0.2}}
\put( 27.00, 15.26){\circle*{0.2}}
\put( 27.05, 15.47){\circle*{0.2}}
\put( 27.11, 15.67){\circle*{0.2}}
\put( 27.18, 15.86){\circle*{0.2}}
\put( 27.25, 16.04){\circle*{0.2}}
\put( 27.33, 16.19){\circle*{0.2}}
\put( 27.42, 16.33){\circle*{0.2}}
\put( 27.52, 16.46){\circle*{0.2}}
\put( 27.63, 16.57){\circle*{0.2}}
\put( 27.74, 16.65){\circle*{0.2}}
\put( 27.86, 16.73){\circle*{0.2}}
\put( 28.00, 16.78){\circle*{0.2}}
\put( 28.14, 16.81){\circle*{0.2}}
\put( 28.29, 16.83){\circle*{0.2}}
\put( 28.45, 16.83){\circle*{0.2}}
\put( 28.62, 16.81){\circle*{0.2}}
\put( 28.80, 16.78){\circle*{0.2}}
\put( 28.98, 16.73){\circle*{0.2}}
\put( 29.17, 16.67){\circle*{0.2}}
\put( 29.37, 16.59){\circle*{0.2}}
\put( 29.57, 16.50){\circle*{0.2}}
\put( 29.78, 16.40){\circle*{0.2}}
\put( 30.00, 16.29){\circle*{0.2}}
\put( 30.22, 16.17){\circle*{0.2}}
\put( 30.44, 16.05){\circle*{0.2}}
\put( 30.66, 15.92){\circle*{0.2}}
\put( 30.89, 15.78){\circle*{0.2}}
\put( 31.12, 15.65){\circle*{0.2}}
\put( 31.35, 15.51){\circle*{0.2}}
\put( 31.57, 15.37){\circle*{0.2}}
\put( 31.80, 15.24){\circle*{0.2}}
\put( 32.02, 15.11){\circle*{0.2}}
\put( 32.25, 14.99){\circle*{0.2}}
\put( 32.46, 14.87){\circle*{0.2}}
\put( 32.68, 14.76){\circle*{0.2}}
\put( 32.88, 14.67){\circle*{0.2}}
\put( 33.09, 14.58){\circle*{0.2}}
\put( 33.28, 14.51){\circle*{0.2}}
\put( 33.47, 14.45){\circle*{0.2}}
\put( 33.65, 14.40){\circle*{0.2}}
\put( 33.83, 14.38){\circle*{0.2}}
\put( 33.99, 14.36){\circle*{0.2}}
\put( 34.15, 14.37){\circle*{0.2}}
\put( 34.30, 14.39){\circle*{0.2}}
\put( 34.44, 14.43){\circle*{0.2}}
\put( 34.57, 14.49){\circle*{0.2}}
\put( 34.69, 14.57){\circle*{0.2}}
\put( 34.80, 14.66){\circle*{0.2}}
\put( 34.90, 14.77){\circle*{0.2}}
\put( 35.00, 14.90){\circle*{0.2}}
\put( 35.09, 15.05){\circle*{0.2}}
\put( 35.16, 15.21){\circle*{0.2}}
\put( 35.24, 15.39){\circle*{0.2}}
\put( 35.30, 15.58){\circle*{0.2}}
\put( 35.36, 15.79){\circle*{0.2}}
\put( 35.41, 16.01){\circle*{0.2}}
\put( 35.45, 16.24){\circle*{0.2}}
\put( 35.49, 16.47){\circle*{0.2}}
\put( 35.53, 16.72){\circle*{0.2}}
\put( 35.56, 16.97){\circle*{0.2}}
\put( 35.59, 17.23){\circle*{0.2}}
\put( 35.62, 17.50){\circle*{0.2}}
\put( 35.65, 17.76){\circle*{0.2}}
\put( 35.67, 18.02){\circle*{0.2}}
\put( 35.70, 18.29){\circle*{0.2}}
\put( 35.73, 18.55){\circle*{0.2}}
\put( 35.76, 18.81){\circle*{0.2}}
\put( 35.80, 19.06){\circle*{0.2}}
\put( 35.83, 19.30){\circle*{0.2}}
\put( 35.88, 19.53){\circle*{0.2}}
\put( 35.92, 19.76){\circle*{0.2}}
\put( 35.98, 19.97){\circle*{0.2}}
\put( 36.04, 20.17){\circle*{0.2}}
\put( 36.11, 20.35){\circle*{0.2}}
\put( 36.18, 20.52){\circle*{0.2}}
\put( 36.26, 20.68){\circle*{0.2}}
\put( 36.35, 20.82){\circle*{0.2}}
\put( 36.45, 20.94){\circle*{0.2}}
\put( 36.56, 21.04){\circle*{0.2}}
\put( 36.68, 21.13){\circle*{0.2}}
\put( 36.80, 21.19){\circle*{0.2}}
\put( 36.94, 21.24){\circle*{0.2}}
\put( 37.08, 21.27){\circle*{0.2}}
\put( 37.23, 21.29){\circle*{0.2}}
\put( 37.39, 21.29){\circle*{0.2}}
\put( 37.56, 21.26){\circle*{0.2}}
\put( 37.74, 21.23){\circle*{0.2}}
\put( 37.93, 21.18){\circle*{0.2}}
\put( 38.12, 21.11){\circle*{0.2}}
\put( 38.32, 21.03){\circle*{0.2}}
\put( 38.52, 20.94){\circle*{0.2}}
\put( 38.74, 20.84){\circle*{0.2}}
\put( 38.95, 20.73){\circle*{0.2}}
\put( 39.17, 20.61){\circle*{0.2}}
\put( 39.39, 20.48){\circle*{0.2}}
\put( 39.62, 20.35){\circle*{0.2}}
\put( 39.85, 20.21){\circle*{0.2}}
\put( 40.07, 20.08){\circle*{0.2}}
}

\newsavebox{\brgluon}
\savebox{\brgluon}(0,0)[tl]{
\put(  0.00,  0.00){\circle*{0.2}}
\put(  0.03, -0.26){\circle*{0.2}}
\put(  0.06, -0.53){\circle*{0.2}}
\put(  0.08, -0.79){\circle*{0.2}}
\put(  0.12, -1.04){\circle*{0.2}}
\put(  0.15, -1.29){\circle*{0.2}}
\put(  0.19, -1.53){\circle*{0.2}}
\put(  0.24, -1.76){\circle*{0.2}}
\put(  0.28, -1.98){\circle*{0.2}}
\put(  0.34, -2.19){\circle*{0.2}}
\put(  0.40, -2.39){\circle*{0.2}}
\put(  0.47, -2.57){\circle*{0.2}}
\put(  0.55, -2.74){\circle*{0.2}}
\put(  0.63, -2.89){\circle*{0.2}}
\put(  0.72, -3.02){\circle*{0.2}}
\put(  0.82, -3.14){\circle*{0.2}}
\put(  0.93, -3.24){\circle*{0.2}}
\put(  1.05, -3.32){\circle*{0.2}}
\put(  1.18, -3.38){\circle*{0.2}}
\put(  1.32, -3.43){\circle*{0.2}}
\put(  1.47, -3.45){\circle*{0.2}}
\put(  1.62, -3.46){\circle*{0.2}}
\put(  1.78, -3.45){\circle*{0.2}}
\put(  1.96, -3.43){\circle*{0.2}}
\put(  2.14, -3.39){\circle*{0.2}}
\put(  2.32, -3.33){\circle*{0.2}}
\put(  2.52, -3.27){\circle*{0.2}}
\put(  2.72, -3.18){\circle*{0.2}}
\put(  2.92, -3.09){\circle*{0.2}}
\put(  3.14, -2.98){\circle*{0.2}}
\put(  3.35, -2.87){\circle*{0.2}}
\put(  3.57, -2.75){\circle*{0.2}}
\put(  3.80, -2.62){\circle*{0.2}}
\put(  4.02, -2.49){\circle*{0.2}}
\put(  4.25, -2.35){\circle*{0.2}}
\put(  4.48, -2.22){\circle*{0.2}}
\put(  4.71, -2.08){\circle*{0.2}}
\put(  4.93, -1.94){\circle*{0.2}}
\put(  5.16, -1.81){\circle*{0.2}}
\put(  5.38, -1.68){\circle*{0.2}}
\put(  5.60, -1.56){\circle*{0.2}}
\put(  5.82, -1.45){\circle*{0.2}}
\put(  6.03, -1.35){\circle*{0.2}}
\put(  6.23, -1.26){\circle*{0.2}}
\put(  6.43, -1.18){\circle*{0.2}}
\put(  6.63, -1.11){\circle*{0.2}}
\put(  6.81, -1.06){\circle*{0.2}}
\put(  6.99, -1.02){\circle*{0.2}}
\put(  7.16, -1.00){\circle*{0.2}}
\put(  7.32, -0.99){\circle*{0.2}}
\put(  7.48, -1.01){\circle*{0.2}}
\put(  7.62, -1.04){\circle*{0.2}}
\put(  7.76, -1.09){\circle*{0.2}}
\put(  7.88, -1.15){\circle*{0.2}}
\put(  8.00, -1.24){\circle*{0.2}}
\put(  8.11, -1.34){\circle*{0.2}}
\put(  8.21, -1.46){\circle*{0.2}}
\put(  8.30, -1.60){\circle*{0.2}}
\put(  8.38, -1.75){\circle*{0.2}}
\put(  8.46, -1.92){\circle*{0.2}}
\put(  8.52, -2.10){\circle*{0.2}}
\put(  8.58, -2.30){\circle*{0.2}}
\put(  8.64, -2.51){\circle*{0.2}}
\put(  8.69, -2.74){\circle*{0.2}}
\put(  8.73, -2.97){\circle*{0.2}}
\put(  8.77, -3.21){\circle*{0.2}}
\put(  8.80, -3.46){\circle*{0.2}}
\put(  8.83, -3.72){\circle*{0.2}}
\put(  8.86, -3.98){\circle*{0.2}}
\put(  8.89, -4.24){\circle*{0.2}}
\put(  8.92, -4.51){\circle*{0.2}}
\put(  8.95, -4.77){\circle*{0.2}}
\put(  8.97, -5.03){\circle*{0.2}}
\put(  9.00, -5.29){\circle*{0.2}}
\put(  9.04, -5.55){\circle*{0.2}}
\put(  9.07, -5.79){\circle*{0.2}}
\put(  9.11, -6.03){\circle*{0.2}}
\put(  9.16, -6.26){\circle*{0.2}}
\put(  9.21, -6.48){\circle*{0.2}}
\put(  9.26, -6.69){\circle*{0.2}}
\put(  9.33, -6.88){\circle*{0.2}}
\put(  9.40, -7.06){\circle*{0.2}}
\put(  9.47, -7.22){\circle*{0.2}}
\put(  9.56, -7.37){\circle*{0.2}}
\put(  9.66, -7.50){\circle*{0.2}}
\put(  9.76, -7.61){\circle*{0.2}}
\put(  9.87, -7.71){\circle*{0.2}}
\put(  9.99, -7.79){\circle*{0.2}}
\put( 10.12, -7.85){\circle*{0.2}}
\put( 10.26, -7.89){\circle*{0.2}}
\put( 10.41, -7.91){\circle*{0.2}}
\put( 10.56, -7.92){\circle*{0.2}}
\put( 10.73, -7.91){\circle*{0.2}}
\put( 10.90, -7.88){\circle*{0.2}}
\put( 11.08, -7.84){\circle*{0.2}}
\put( 11.27, -7.78){\circle*{0.2}}
\put( 11.47, -7.71){\circle*{0.2}}
\put( 11.67, -7.62){\circle*{0.2}}
\put( 11.88, -7.53){\circle*{0.2}}
\put( 12.09, -7.42){\circle*{0.2}}
\put( 12.31, -7.30){\circle*{0.2}}
\put( 12.53, -7.18){\circle*{0.2}}
\put( 12.75, -7.05){\circle*{0.2}}
\put( 12.98, -6.92){\circle*{0.2}}
\put( 13.21, -6.78){\circle*{0.2}}
\put( 13.43, -6.65){\circle*{0.2}}
\put( 13.66, -6.51){\circle*{0.2}}
\put( 13.89, -6.38){\circle*{0.2}}
\put( 14.11, -6.24){\circle*{0.2}}
\put( 14.34, -6.12){\circle*{0.2}}
\put( 14.56, -6.00){\circle*{0.2}}
\put( 14.77, -5.89){\circle*{0.2}}
\put( 14.98, -5.79){\circle*{0.2}}
\put( 15.18, -5.70){\circle*{0.2}}
\put( 15.38, -5.62){\circle*{0.2}}
\put( 15.58, -5.56){\circle*{0.2}}
\put( 15.76, -5.51){\circle*{0.2}}
\put( 15.94, -5.47){\circle*{0.2}}
\put( 16.11, -5.45){\circle*{0.2}}
\put( 16.27, -5.45){\circle*{0.2}}
\put( 16.42, -5.47){\circle*{0.2}}
\put( 16.56, -5.50){\circle*{0.2}}
\put( 16.69, -5.55){\circle*{0.2}}
\put( 16.82, -5.62){\circle*{0.2}}
\put( 16.93, -5.71){\circle*{0.2}}
\put( 17.04, -5.82){\circle*{0.2}}
\put( 17.14, -5.94){\circle*{0.2}}
\put( 17.23, -6.08){\circle*{0.2}}
\put( 17.31, -6.24){\circle*{0.2}}
\put( 17.38, -6.41){\circle*{0.2}}
\put( 17.45, -6.60){\circle*{0.2}}
\put( 17.51, -6.80){\circle*{0.2}}
\put( 17.56, -7.01){\circle*{0.2}}
\put( 17.61, -7.24){\circle*{0.2}}
\put( 17.65, -7.47){\circle*{0.2}}
\put( 17.69, -7.71){\circle*{0.2}}
\put( 17.72, -7.97){\circle*{0.2}}
\put( 17.75, -8.22){\circle*{0.2}}
\put( 17.78, -8.48){\circle*{0.2}}
\put( 17.81, -8.75){\circle*{0.2}}
\put( 17.84, -9.01){\circle*{0.2}}
\put( 17.86, -9.28){\circle*{0.2}}
\put( 17.89, -9.54){\circle*{0.2}}
\put( 17.92, -9.80){\circle*{0.2}}
\put( 17.96,-10.05){\circle*{0.2}}
\put( 17.99,-10.30){\circle*{0.2}}
\put( 18.03,-10.53){\circle*{0.2}}
\put( 18.08,-10.76){\circle*{0.2}}
\put( 18.13,-10.98){\circle*{0.2}}
\put( 18.19,-11.18){\circle*{0.2}}
\put( 18.25,-11.37){\circle*{0.2}}
\put( 18.32,-11.55){\circle*{0.2}}
\put( 18.40,-11.71){\circle*{0.2}}
\put( 18.49,-11.85){\circle*{0.2}}
\put( 18.59,-11.98){\circle*{0.2}}
\put( 18.69,-12.09){\circle*{0.2}}
\put( 18.80,-12.18){\circle*{0.2}}
\put( 18.93,-12.26){\circle*{0.2}}
\put( 19.06,-12.31){\circle*{0.2}}
\put( 19.20,-12.35){\circle*{0.2}}
\put( 19.35,-12.37){\circle*{0.2}}
\put( 19.51,-12.38){\circle*{0.2}}
\put( 19.67,-12.36){\circle*{0.2}}
\put( 19.85,-12.33){\circle*{0.2}}
\put( 20.03,-12.28){\circle*{0.2}}
\put( 20.22,-12.22){\circle*{0.2}}
\put( 20.42,-12.15){\circle*{0.2}}
\put( 20.62,-12.06){\circle*{0.2}}
\put( 20.83,-11.96){\circle*{0.2}}
\put( 21.04,-11.86){\circle*{0.2}}
\put( 21.26,-11.74){\circle*{0.2}}
\put( 21.48,-11.61){\circle*{0.2}}
\put( 21.71,-11.49){\circle*{0.2}}
\put( 21.93,-11.35){\circle*{0.2}}
\put( 22.16,-11.21){\circle*{0.2}}
\put( 22.39,-11.08){\circle*{0.2}}
\put( 22.62,-10.94){\circle*{0.2}}
\put( 22.84,-10.81){\circle*{0.2}}
\put( 23.07,-10.68){\circle*{0.2}}
\put( 23.29,-10.55){\circle*{0.2}}
\put( 23.51,-10.43){\circle*{0.2}}
\put( 23.72,-10.33){\circle*{0.2}}
\put( 23.93,-10.23){\circle*{0.2}}
\put( 24.14,-10.14){\circle*{0.2}}
\put( 24.33,-10.06){\circle*{0.2}}
\put( 24.52,-10.00){\circle*{0.2}}
\put( 24.71, -9.96){\circle*{0.2}}
\put( 24.88, -9.92){\circle*{0.2}}
\put( 25.05, -9.91){\circle*{0.2}}
\put( 25.21, -9.91){\circle*{0.2}}
\put( 25.36, -9.93){\circle*{0.2}}
\put( 25.50, -9.97){\circle*{0.2}}
\put( 25.63,-10.02){\circle*{0.2}}
\put( 25.75,-10.09){\circle*{0.2}}
\put( 25.87,-10.19){\circle*{0.2}}
\put( 25.97,-10.29){\circle*{0.2}}
\put( 26.07,-10.42){\circle*{0.2}}
\put( 26.16,-10.56){\circle*{0.2}}
\put( 26.24,-10.72){\circle*{0.2}}
\put( 26.31,-10.90){\circle*{0.2}}
\put( 26.37,-11.09){\circle*{0.2}}
\put( 26.43,-11.29){\circle*{0.2}}
\put( 26.48,-11.51){\circle*{0.2}}
\put( 26.53,-11.73){\circle*{0.2}}
\put( 26.57,-11.97){\circle*{0.2}}
\put( 26.61,-12.22){\circle*{0.2}}
\put( 26.64,-12.47){\circle*{0.2}}
\put( 26.67,-12.73){\circle*{0.2}}
\put( 26.70,-12.99){\circle*{0.2}}
\put( 26.73,-13.25){\circle*{0.2}}
\put( 26.76,-13.52){\circle*{0.2}}
\put( 26.78,-13.78){\circle*{0.2}}
\put( 26.81,-14.04){\circle*{0.2}}
\put( 26.84,-14.30){\circle*{0.2}}
\put( 26.88,-14.55){\circle*{0.2}}
\put( 26.91,-14.80){\circle*{0.2}}
\put( 26.96,-15.03){\circle*{0.2}}
\put( 27.00,-15.26){\circle*{0.2}}
\put( 27.05,-15.47){\circle*{0.2}}
\put( 27.11,-15.67){\circle*{0.2}}
\put( 27.18,-15.86){\circle*{0.2}}
\put( 27.25,-16.04){\circle*{0.2}}
\put( 27.33,-16.19){\circle*{0.2}}
\put( 27.42,-16.33){\circle*{0.2}}
\put( 27.52,-16.46){\circle*{0.2}}
\put( 27.63,-16.57){\circle*{0.2}}
\put( 27.74,-16.65){\circle*{0.2}}
\put( 27.86,-16.73){\circle*{0.2}}
\put( 28.00,-16.78){\circle*{0.2}}
\put( 28.14,-16.81){\circle*{0.2}}
\put( 28.29,-16.83){\circle*{0.2}}
\put( 28.45,-16.83){\circle*{0.2}}
\put( 28.62,-16.81){\circle*{0.2}}
\put( 28.80,-16.78){\circle*{0.2}}
\put( 28.98,-16.73){\circle*{0.2}}
\put( 29.17,-16.67){\circle*{0.2}}
\put( 29.37,-16.59){\circle*{0.2}}
\put( 29.57,-16.50){\circle*{0.2}}
\put( 29.78,-16.40){\circle*{0.2}}
\put( 30.00,-16.29){\circle*{0.2}}
\put( 30.22,-16.17){\circle*{0.2}}
\put( 30.44,-16.05){\circle*{0.2}}
\put( 30.66,-15.92){\circle*{0.2}}
\put( 30.89,-15.78){\circle*{0.2}}
\put( 31.12,-15.65){\circle*{0.2}}
\put( 31.35,-15.51){\circle*{0.2}}
\put( 31.57,-15.37){\circle*{0.2}}
\put( 31.80,-15.24){\circle*{0.2}}
\put( 32.02,-15.11){\circle*{0.2}}
\put( 32.25,-14.99){\circle*{0.2}}
\put( 32.46,-14.87){\circle*{0.2}}
\put( 32.68,-14.76){\circle*{0.2}}
\put( 32.88,-14.67){\circle*{0.2}}
\put( 33.09,-14.58){\circle*{0.2}}
\put( 33.28,-14.51){\circle*{0.2}}
\put( 33.47,-14.45){\circle*{0.2}}
\put( 33.65,-14.40){\circle*{0.2}}
\put( 33.83,-14.38){\circle*{0.2}}
\put( 33.99,-14.36){\circle*{0.2}}
\put( 34.15,-14.37){\circle*{0.2}}
\put( 34.30,-14.39){\circle*{0.2}}
\put( 34.44,-14.43){\circle*{0.2}}
\put( 34.57,-14.49){\circle*{0.2}}
\put( 34.69,-14.57){\circle*{0.2}}
\put( 34.80,-14.66){\circle*{0.2}}
\put( 34.90,-14.77){\circle*{0.2}}
\put( 35.00,-14.90){\circle*{0.2}}
\put( 35.09,-15.05){\circle*{0.2}}
\put( 35.16,-15.21){\circle*{0.2}}
\put( 35.24,-15.39){\circle*{0.2}}
\put( 35.30,-15.58){\circle*{0.2}}
\put( 35.36,-15.79){\circle*{0.2}}
\put( 35.41,-16.01){\circle*{0.2}}
\put( 35.45,-16.24){\circle*{0.2}}
\put( 35.49,-16.47){\circle*{0.2}}
\put( 35.53,-16.72){\circle*{0.2}}
\put( 35.56,-16.97){\circle*{0.2}}
\put( 35.59,-17.23){\circle*{0.2}}
\put( 35.62,-17.50){\circle*{0.2}}
\put( 35.65,-17.76){\circle*{0.2}}
\put( 35.67,-18.02){\circle*{0.2}}
\put( 35.70,-18.29){\circle*{0.2}}
\put( 35.73,-18.55){\circle*{0.2}}
\put( 35.76,-18.81){\circle*{0.2}}
\put( 35.80,-19.06){\circle*{0.2}}
\put( 35.83,-19.30){\circle*{0.2}}
\put( 35.88,-19.53){\circle*{0.2}}
\put( 35.92,-19.76){\circle*{0.2}}
\put( 35.98,-19.97){\circle*{0.2}}
\put( 36.04,-20.17){\circle*{0.2}}
\put( 36.11,-20.35){\circle*{0.2}}
\put( 36.18,-20.52){\circle*{0.2}}
\put( 36.26,-20.68){\circle*{0.2}}
\put( 36.35,-20.82){\circle*{0.2}}
\put( 36.45,-20.94){\circle*{0.2}}
\put( 36.56,-21.04){\circle*{0.2}}
\put( 36.68,-21.13){\circle*{0.2}}
\put( 36.80,-21.19){\circle*{0.2}}
\put( 36.94,-21.24){\circle*{0.2}}
\put( 37.08,-21.27){\circle*{0.2}}
\put( 37.23,-21.29){\circle*{0.2}}
\put( 37.39,-21.29){\circle*{0.2}}
\put( 37.56,-21.26){\circle*{0.2}}
\put( 37.74,-21.23){\circle*{0.2}}
\put( 37.93,-21.18){\circle*{0.2}}
\put( 38.12,-21.11){\circle*{0.2}}
\put( 38.32,-21.03){\circle*{0.2}}
\put( 38.52,-20.94){\circle*{0.2}}
\put( 38.74,-20.84){\circle*{0.2}}
\put( 38.95,-20.73){\circle*{0.2}}
\put( 39.17,-20.61){\circle*{0.2}}
\put( 39.39,-20.48){\circle*{0.2}}
\put( 39.62,-20.35){\circle*{0.2}}
\put( 39.85,-20.21){\circle*{0.2}}
\put( 40.07,-20.08){\circle*{0.2}}
}

\newsavebox{\topgluon}
\savebox{\topgluon}(0,0)[c]{
\put(  0.00,  2.50){\circle*{0.2}}
\put(  0.13,  2.49){\circle*{0.2}}
\put(  0.25,  2.47){\circle*{0.2}}
\put(  0.38,  2.43){\circle*{0.2}}
\put(  0.51,  2.37){\circle*{0.2}}
\put(  0.64,  2.30){\circle*{0.2}}
\put(  0.76,  2.22){\circle*{0.2}}
\put(  0.89,  2.12){\circle*{0.2}}
\put(  1.02,  2.01){\circle*{0.2}}
\put(  1.15,  1.88){\circle*{0.2}}
\put(  1.27,  1.74){\circle*{0.2}}
\put(  1.40,  1.59){\circle*{0.2}}
\put(  1.53,  1.43){\circle*{0.2}}
\put(  1.66,  1.27){\circle*{0.2}}
\put(  1.78,  1.09){\circle*{0.2}}
\put(  1.91,  0.91){\circle*{0.2}}
\put(  2.04,  0.72){\circle*{0.2}}
\put(  2.17,  0.52){\circle*{0.2}}
\put(  2.29,  0.33){\circle*{0.2}}
\put(  2.42,  0.13){\circle*{0.2}}
\put(  2.55, -0.07){\circle*{0.2}}
\put(  2.68, -0.27){\circle*{0.2}}
\put(  2.80, -0.47){\circle*{0.2}}
\put(  2.93, -0.66){\circle*{0.2}}
\put(  3.06, -0.86){\circle*{0.2}}
\put(  3.18, -1.04){\circle*{0.2}}
\put(  3.31, -1.22){\circle*{0.2}}
\put(  3.44, -1.39){\circle*{0.2}}
\put(  3.57, -1.55){\circle*{0.2}}
\put(  3.69, -1.70){\circle*{0.2}}
\put(  3.82, -1.84){\circle*{0.2}}
\put(  3.95, -1.97){\circle*{0.2}}
\put(  4.08, -2.09){\circle*{0.2}}
\put(  4.20, -2.19){\circle*{0.2}}
\put(  4.33, -2.28){\circle*{0.2}}
\put(  4.46, -2.36){\circle*{0.2}}
\put(  4.59, -2.41){\circle*{0.2}}
\put(  4.71, -2.46){\circle*{0.2}}
\put(  4.84, -2.49){\circle*{0.2}}
\put(  4.97, -2.50){\circle*{0.2}}
\put(  5.10, -2.50){\circle*{0.2}}
\put(  5.22, -2.48){\circle*{0.2}}
\put(  5.35, -2.44){\circle*{0.2}}
\put(  5.48, -2.39){\circle*{0.2}}
\put(  5.61, -2.32){\circle*{0.2}}
\put(  5.73, -2.24){\circle*{0.2}}
\put(  5.86, -2.15){\circle*{0.2}}
\put(  5.99, -2.04){\circle*{0.2}}
\put(  6.11, -1.91){\circle*{0.2}}
\put(  6.24, -1.78){\circle*{0.2}}
\put(  6.37, -1.63){\circle*{0.2}}
\put(  6.50, -1.48){\circle*{0.2}}
\put(  6.62, -1.31){\circle*{0.2}}
\put(  6.75, -1.14){\circle*{0.2}}
\put(  6.88, -0.96){\circle*{0.2}}
\put(  7.01, -0.77){\circle*{0.2}}
\put(  7.13, -0.58){\circle*{0.2}}
\put(  7.26, -0.38){\circle*{0.2}}
\put(  7.39, -0.18){\circle*{0.2}}
\put(  7.52,  0.02){\circle*{0.2}}
\put(  7.64,  0.22){\circle*{0.2}}
\put(  7.77,  0.42){\circle*{0.2}}
\put(  7.90,  0.61){\circle*{0.2}}
\put(  8.03,  0.80){\circle*{0.2}}
\put(  8.15,  0.99){\circle*{0.2}}
\put(  8.28,  1.17){\circle*{0.2}}
\put(  8.41,  1.34){\circle*{0.2}}
\put(  8.54,  1.51){\circle*{0.2}}
\put(  8.66,  1.66){\circle*{0.2}}
\put(  8.79,  1.81){\circle*{0.2}}
\put(  8.92,  1.94){\circle*{0.2}}
\put(  9.04,  2.06){\circle*{0.2}}
\put(  9.17,  2.17){\circle*{0.2}}
\put(  9.30,  2.26){\circle*{0.2}}
\put(  9.43,  2.34){\circle*{0.2}}
\put(  9.55,  2.40){\circle*{0.2}}
\put(  9.68,  2.45){\circle*{0.2}}
\put(  9.81,  2.48){\circle*{0.2}}
\put(  9.94,  2.50){\circle*{0.2}}
\put( 10.06,  2.50){\circle*{0.2}}
\put( 10.19,  2.48){\circle*{0.2}}
\put( 10.32,  2.45){\circle*{0.2}}
\put( 10.45,  2.40){\circle*{0.2}}
\put( 10.57,  2.34){\circle*{0.2}}
\put( 10.70,  2.27){\circle*{0.2}}
\put( 10.83,  2.17){\circle*{0.2}}
\put( 10.96,  2.07){\circle*{0.2}}
\put( 11.08,  1.95){\circle*{0.2}}
\put( 11.21,  1.82){\circle*{0.2}}
\put( 11.34,  1.67){\circle*{0.2}}
\put( 11.46,  1.52){\circle*{0.2}}
\put( 11.59,  1.36){\circle*{0.2}}
\put( 11.72,  1.19){\circle*{0.2}}
\put( 11.85,  1.01){\circle*{0.2}}
\put( 11.97,  0.82){\circle*{0.2}}
\put( 12.10,  0.63){\circle*{0.2}}
\put( 12.23,  0.43){\circle*{0.2}}
\put( 12.36,  0.23){\circle*{0.2}}
\put( 12.48,  0.03){\circle*{0.2}}
\put( 12.61, -0.16){\circle*{0.2}}
\put( 12.74, -0.36){\circle*{0.2}}
\put( 12.87, -0.56){\circle*{0.2}}
\put( 12.99, -0.75){\circle*{0.2}}
\put( 13.12, -0.94){\circle*{0.2}}
\put( 13.25, -1.12){\circle*{0.2}}
\put( 13.38, -1.30){\circle*{0.2}}
\put( 13.50, -1.46){\circle*{0.2}}
\put( 13.63, -1.62){\circle*{0.2}}
\put( 13.76, -1.77){\circle*{0.2}}
\put( 13.89, -1.90){\circle*{0.2}}
\put( 14.01, -2.03){\circle*{0.2}}
\put( 14.14, -2.14){\circle*{0.2}}
\put( 14.27, -2.23){\circle*{0.2}}
\put( 14.39, -2.32){\circle*{0.2}}
\put( 14.52, -2.38){\circle*{0.2}}
\put( 14.65, -2.44){\circle*{0.2}}
\put( 14.78, -2.47){\circle*{0.2}}
\put( 14.90, -2.49){\circle*{0.2}}
\put( 15.03, -2.50){\circle*{0.2}}
\put( 15.16, -2.49){\circle*{0.2}}
\put( 15.29, -2.46){\circle*{0.2}}
\put( 15.41, -2.42){\circle*{0.2}}
\put( 15.54, -2.36){\circle*{0.2}}
\put( 15.67, -2.29){\circle*{0.2}}
\put( 15.80, -2.20){\circle*{0.2}}
\put( 15.92, -2.10){\circle*{0.2}}
\put( 16.05, -1.98){\circle*{0.2}}
\put( 16.18, -1.85){\circle*{0.2}}
\put( 16.31, -1.71){\circle*{0.2}}
\put( 16.43, -1.56){\circle*{0.2}}
\put( 16.56, -1.40){\circle*{0.2}}
\put( 16.69, -1.23){\circle*{0.2}}
\put( 16.82, -1.05){\circle*{0.2}}
\put( 16.94, -0.87){\circle*{0.2}}
\put( 17.07, -0.68){\circle*{0.2}}
\put( 17.20, -0.49){\circle*{0.2}}
\put( 17.32, -0.29){\circle*{0.2}}
\put( 17.45, -0.09){\circle*{0.2}}
\put( 17.58,  0.11){\circle*{0.2}}
\put( 17.71,  0.31){\circle*{0.2}}
\put( 17.83,  0.51){\circle*{0.2}}
\put( 17.96,  0.70){\circle*{0.2}}
\put( 18.09,  0.89){\circle*{0.2}}
\put( 18.22,  1.07){\circle*{0.2}}
\put( 18.34,  1.25){\circle*{0.2}}
\put( 18.47,  1.42){\circle*{0.2}}
\put( 18.60,  1.58){\circle*{0.2}}
\put( 18.73,  1.73){\circle*{0.2}}
\put( 18.85,  1.87){\circle*{0.2}}
\put( 18.98,  2.00){\circle*{0.2}}
\put( 19.11,  2.11){\circle*{0.2}}
\put( 19.24,  2.21){\circle*{0.2}}
\put( 19.36,  2.30){\circle*{0.2}}
\put( 19.49,  2.37){\circle*{0.2}}
\put( 19.62,  2.42){\circle*{0.2}}
\put( 19.75,  2.47){\circle*{0.2}}
\put( 19.87,  2.49){\circle*{0.2}}
}

\newsavebox{\bottomgluon}
\savebox{\bottomgluon}(0,0)[c]{
\put(  0.00, -2.50){\circle*{0.2}}
\put(  0.13, -2.49){\circle*{0.2}}
\put(  0.25, -2.47){\circle*{0.2}}
\put(  0.38, -2.43){\circle*{0.2}}
\put(  0.51, -2.37){\circle*{0.2}}
\put(  0.64, -2.30){\circle*{0.2}}
\put(  0.76, -2.22){\circle*{0.2}}
\put(  0.89, -2.12){\circle*{0.2}}
\put(  1.02, -2.01){\circle*{0.2}}
\put(  1.15, -1.88){\circle*{0.2}}
\put(  1.27, -1.74){\circle*{0.2}}
\put(  1.40, -1.59){\circle*{0.2}}
\put(  1.53, -1.43){\circle*{0.2}}
\put(  1.66, -1.27){\circle*{0.2}}
\put(  1.78, -1.09){\circle*{0.2}}
\put(  1.91, -0.91){\circle*{0.2}}
\put(  2.04, -0.72){\circle*{0.2}}
\put(  2.17, -0.52){\circle*{0.2}}
\put(  2.29, -0.33){\circle*{0.2}}
\put(  2.42, -0.13){\circle*{0.2}}
\put(  2.55,  0.07){\circle*{0.2}}
\put(  2.68,  0.27){\circle*{0.2}}
\put(  2.80,  0.47){\circle*{0.2}}
\put(  2.93,  0.66){\circle*{0.2}}
\put(  3.06,  0.86){\circle*{0.2}}
\put(  3.18,  1.04){\circle*{0.2}}
\put(  3.31,  1.22){\circle*{0.2}}
\put(  3.44,  1.39){\circle*{0.2}}
\put(  3.57,  1.55){\circle*{0.2}}
\put(  3.69,  1.70){\circle*{0.2}}
\put(  3.82,  1.84){\circle*{0.2}}
\put(  3.95,  1.97){\circle*{0.2}}
\put(  4.08,  2.09){\circle*{0.2}}
\put(  4.20,  2.19){\circle*{0.2}}
\put(  4.33,  2.28){\circle*{0.2}}
\put(  4.46,  2.36){\circle*{0.2}}
\put(  4.59,  2.41){\circle*{0.2}}
\put(  4.71,  2.46){\circle*{0.2}}
\put(  4.84,  2.49){\circle*{0.2}}
\put(  4.97,  2.50){\circle*{0.2}}
\put(  5.10,  2.50){\circle*{0.2}}
\put(  5.22,  2.48){\circle*{0.2}}
\put(  5.35,  2.44){\circle*{0.2}}
\put(  5.48,  2.39){\circle*{0.2}}
\put(  5.61,  2.32){\circle*{0.2}}
\put(  5.73,  2.24){\circle*{0.2}}
\put(  5.86,  2.15){\circle*{0.2}}
\put(  5.99,  2.04){\circle*{0.2}}
\put(  6.11,  1.91){\circle*{0.2}}
\put(  6.24,  1.78){\circle*{0.2}}
\put(  6.37,  1.63){\circle*{0.2}}
\put(  6.50,  1.48){\circle*{0.2}}
\put(  6.62,  1.31){\circle*{0.2}}
\put(  6.75,  1.14){\circle*{0.2}}
\put(  6.88,  0.96){\circle*{0.2}}
\put(  7.01,  0.77){\circle*{0.2}}
\put(  7.13,  0.58){\circle*{0.2}}
\put(  7.26,  0.38){\circle*{0.2}}
\put(  7.39,  0.18){\circle*{0.2}}
\put(  7.52, -0.02){\circle*{0.2}}
\put(  7.64, -0.22){\circle*{0.2}}
\put(  7.77, -0.42){\circle*{0.2}}
\put(  7.90, -0.61){\circle*{0.2}}
\put(  8.03, -0.80){\circle*{0.2}}
\put(  8.15, -0.99){\circle*{0.2}}
\put(  8.28, -1.17){\circle*{0.2}}
\put(  8.41, -1.34){\circle*{0.2}}
\put(  8.54, -1.51){\circle*{0.2}}
\put(  8.66, -1.66){\circle*{0.2}}
\put(  8.79, -1.81){\circle*{0.2}}
\put(  8.92, -1.94){\circle*{0.2}}
\put(  9.04, -2.06){\circle*{0.2}}
\put(  9.17, -2.17){\circle*{0.2}}
\put(  9.30, -2.26){\circle*{0.2}}
\put(  9.43, -2.34){\circle*{0.2}}
\put(  9.55, -2.40){\circle*{0.2}}
\put(  9.68, -2.45){\circle*{0.2}}
\put(  9.81, -2.48){\circle*{0.2}}
\put(  9.94, -2.50){\circle*{0.2}}
\put( 10.06, -2.50){\circle*{0.2}}
\put( 10.19, -2.48){\circle*{0.2}}
\put( 10.32, -2.45){\circle*{0.2}}
\put( 10.45, -2.40){\circle*{0.2}}
\put( 10.57, -2.34){\circle*{0.2}}
\put( 10.70, -2.27){\circle*{0.2}}
\put( 10.83, -2.17){\circle*{0.2}}
\put( 10.96, -2.07){\circle*{0.2}}
\put( 11.08, -1.95){\circle*{0.2}}
\put( 11.21, -1.82){\circle*{0.2}}
\put( 11.34, -1.67){\circle*{0.2}}
\put( 11.46, -1.52){\circle*{0.2}}
\put( 11.59, -1.36){\circle*{0.2}}
\put( 11.72, -1.19){\circle*{0.2}}
\put( 11.85, -1.01){\circle*{0.2}}
\put( 11.97, -0.82){\circle*{0.2}}
\put( 12.10, -0.63){\circle*{0.2}}
\put( 12.23, -0.43){\circle*{0.2}}
\put( 12.36, -0.23){\circle*{0.2}}
\put( 12.48, -0.03){\circle*{0.2}}
\put( 12.61,  0.16){\circle*{0.2}}
\put( 12.74,  0.36){\circle*{0.2}}
\put( 12.87,  0.56){\circle*{0.2}}
\put( 12.99,  0.75){\circle*{0.2}}
\put( 13.12,  0.94){\circle*{0.2}}
\put( 13.25,  1.12){\circle*{0.2}}
\put( 13.38,  1.30){\circle*{0.2}}
\put( 13.50,  1.46){\circle*{0.2}}
\put( 13.63,  1.62){\circle*{0.2}}
\put( 13.76,  1.77){\circle*{0.2}}
\put( 13.89,  1.90){\circle*{0.2}}
\put( 14.01,  2.03){\circle*{0.2}}
\put( 14.14,  2.14){\circle*{0.2}}
\put( 14.27,  2.23){\circle*{0.2}}
\put( 14.39,  2.32){\circle*{0.2}}
\put( 14.52,  2.38){\circle*{0.2}}
\put( 14.65,  2.44){\circle*{0.2}}
\put( 14.78,  2.47){\circle*{0.2}}
\put( 14.90,  2.49){\circle*{0.2}}
\put( 15.03,  2.50){\circle*{0.2}}
\put( 15.16,  2.49){\circle*{0.2}}
\put( 15.29,  2.46){\circle*{0.2}}
\put( 15.41,  2.42){\circle*{0.2}}
\put( 15.54,  2.36){\circle*{0.2}}
\put( 15.67,  2.29){\circle*{0.2}}
\put( 15.80,  2.20){\circle*{0.2}}
\put( 15.92,  2.10){\circle*{0.2}}
\put( 16.05,  1.98){\circle*{0.2}}
\put( 16.18,  1.85){\circle*{0.2}}
\put( 16.31,  1.71){\circle*{0.2}}
\put( 16.43,  1.56){\circle*{0.2}}
\put( 16.56,  1.40){\circle*{0.2}}
\put( 16.69,  1.23){\circle*{0.2}}
\put( 16.82,  1.05){\circle*{0.2}}
\put( 16.94,  0.87){\circle*{0.2}}
\put( 17.07,  0.68){\circle*{0.2}}
\put( 17.20,  0.49){\circle*{0.2}}
\put( 17.32,  0.29){\circle*{0.2}}
\put( 17.45,  0.09){\circle*{0.2}}
\put( 17.58, -0.11){\circle*{0.2}}
\put( 17.71, -0.31){\circle*{0.2}}
\put( 17.83, -0.51){\circle*{0.2}}
\put( 17.96, -0.70){\circle*{0.2}}
\put( 18.09, -0.89){\circle*{0.2}}
\put( 18.22, -1.07){\circle*{0.2}}
\put( 18.34, -1.25){\circle*{0.2}}
\put( 18.47, -1.42){\circle*{0.2}}
\put( 18.60, -1.58){\circle*{0.2}}
\put( 18.73, -1.73){\circle*{0.2}}
\put( 18.85, -1.87){\circle*{0.2}}
\put( 18.98, -2.00){\circle*{0.2}}
\put( 19.11, -2.11){\circle*{0.2}}
\put( 19.24, -2.21){\circle*{0.2}}
\put( 19.36, -2.30){\circle*{0.2}}
\put( 19.49, -2.37){\circle*{0.2}}
\put( 19.62, -2.42){\circle*{0.2}}
\put( 19.75, -2.47){\circle*{0.2}}
\put( 19.87, -2.49){\circle*{0.2}}
}

\begin{document}

\begin{titlepage}

\begin{center}

{\Large Centre de Physique Th\'eorique - CNRS - Luminy, Case 907}

{\Large F-13288 Marseille Cedex 9, France}

\vspace{3 cm}

{\Large {\bf UNIVERSALITY OF THE PINCH TECHNIQUE
GAUGE BOSON SELF-ENERGIES}}

\vspace{0.3 cm}

{\bf N. Jay Watson}\footnote{email: watson@cptsu4.univ-mrs.fr}

\vspace{2.5 cm}

{\bf Abstract}

\end{center}

It is shown how the S-matrix pinch technique may be extended to the
cases of external scalar and gauge boson fields. Using this
extension, the universality of the pinch technique
gauge-independent one-loop gauge boson self-energy
in a general unbroken SU($N$) gauge theory is demonstrated explicitly.

\vspace{2 cm}

\noindent Key-Words: pinch technique, self-energy

\bigskip

\noindent Number of figures: 3

\bigskip

\noindent December 1994

\noindent CPT-94/P.3133

\bigskip

\noindent anonymous ftp or gopher: cpt.univ-mrs.fr

\end{titlepage}

{\bf 1.} The pinch technique (PT) is a well-defined algorithm for
the rearrangement of contributions from the conventional
gauge-dependent $n$-point functions occurring
in one-loop processes to construct gauge-independent one-loop
self-energy-like, vertex-like and box-like functions
in non-abelian gauge theories.
First introduced by Cornwall
\cite{cornwall1}--\cite{cornwall3}
and further developed by Cornwall and Papavassiliou
\cite{cornwall4}\cite{papav1} in the context of QCD,
the original motivation for the PT
was to enable gauge-independent truncation schemes for non-perturbative
approaches to QCD involving Schwinger-Dyson equations.
Since then, the PT
has also been applied to spontaneously broken gauge theories.
The first such application was made by Papavassiliou \cite{papav2}
in the context of a simplified Georgi-Glashow model.
More recently, the PT has been used
by Degrassi and Sirlin \cite{degrassi1}
in the electroweak sector of the Standard Model
to derive explicitly $\xi$-independent electroweak gauge boson
self-energies in the class of renormalizable $R_{\xi}$ gauges.
Subsequently, there have been various applications in electroweak
phenomenology \cite{degrassi2}--\cite{papav6}.

The PT is based on the observation that in a non-abelian gauge theory,
one-loop diagrams which appear to give only vertex or box corrections
to tree level processes in fact implicitly contain propagator-like
components. This property is a fundamental consequence of the
underlying non-abelian gauge symmetry of the theory, be it unbroken or
broken. For example, in the four-fermion process
$\psi_{i}(p)+\psi_{i'}(p')\rightarrow\psi_{j}(p+q)+\psi_{j'}(p'-q)$,
the conventional gauge boson one-loop two-point function
contribution is as shown schematically in
fig.\ 1(a). This conventional two-point contribution is gauge-dependent.
However, at the one-loop level there also occur the diagrams shown
in figs.\ 1(b), (d) and (f) which implicitly contain
propagator-like contributions. These
propagator-like contributions occur when, in the external fermion
lines, the vector--fermion--fermion interaction vertices coincide.
These are the ``pinch parts'' of the diagrams, shown in
figs.\ 1(c), (e) and (g). In a conventional
$R_{\xi}$ gauge, the $\xi$-dependence of these pinch parts is such as
to cancel exactly that of the conventional two-point
function. The PT gauge-independent one-loop gauge
boson self-energy is therefore constructed
by adding to the conventional self-energy in fig.\ 1(a) the
propagator-like contributions figs.\ 1(c), (e) and (g)
extracted from the conventional vertex
and box functions.
The resulting function is illustrated schematically in fig.\ 1(h).
The PT gauge-independent
one-loop vector--fermion--fermion vertex and four-fermion box functions
are then the conventional functions with the
pinch part contributions subtracted.

There are three principle formulations of the PT: the intrinsic and
the S-matrix formulations of Cornwall and Papavassiliou, and the
current algebra formulation of Degrassi and Sirlin. The S-matrix
and current algebra versions are explicitly formulated in the context
of one-loop processes involving on-shell external fermions, while
the intrinsic PT, although it avoids the explicit embedding in
S-matrix elements,
starts implicitly from the consideration of fermionic processes.
In order however
for the self-energies constructed in the PT to
be intrinsic properties of the gauge bosons, and so
to be physically meaningful, clearly a necessary
condition is that they should be
universal, i.e.\ independent of whether the external fields in the
processes from which the self-energies are constructed are fermions,
scalars or gauge bosons. This necessary condition is in addition to
that of gauge independence. It has been shown
by Degrassi and Sirlin \cite{degrassi1} using an indirect method,
still ultimately involving external fermions,
that the same PT electroweak self-energies are obtained
from the process $e^{+}e^{-} \rightarrow W^{+}W^{-}$ as from
$e^{+}e^{-} \rightarrow e^{+}e^{-}$.
However, a general and direct demonstration of the PT
self-energies' universality is lacking.

In this letter, it is shown how the S-matrix
PT may be extended directly to the cases of external
scalar and gauge boson fields. Using this extension, the
universality of the PT gauge-independent one-loop gauge boson
self-energy in a
general unbroken SU($N$) gauge theory is demonstrated explicitly.

\vspace{10pt}

{\bf 2.} We consider an unbroken SU($N$) gauge theory coupled
both to fermions of mass $m$ in a multiplet $\Psi$ with representation
matices $T^{a}$ and scalars of mass
$M$ $(M^{2} > 0)$ in a complex multiplet $\Phi$
with representation matrices $R^{a}$:
\bea\label{Lcl}
{\cal L}_{\rm cl}
&=&
-\textstyle{\frac{1}{4}}
(\partial_{\mu}A_{\nu}^{a} - \partial_{\nu}A_{\mu}^{a}
+ gf^{abc}A_{\mu}^{b}A_{\nu}^{c})
(\partial^{\mu}A^{a\nu} - \partial^{\nu}A^{a\mu} +
gf^{ade}A^{d\mu}A^{e\nu}) \nonumber \\
& &
+ \overline{\Psi}[i\gamma^{\mu}(\partial_{\mu} - igA_{\mu}^{a}T^{a})
- m]\Psi\nonumber\\
& &
+ [(\partial_{\mu} - igA_{\mu}^{a}R^{a})\Phi]^{\dagger}
[(\partial^{\mu} - igA^{b\mu}R^{b})\Phi]
- M^{2}\Phi^{\dagger}\Phi
- \textstyle{\frac{1}{4!}}\lambda (\Phi^{\dagger}\Phi)^{2}.
\eea
The Feynman rules for the interaction of the gauge boson
$A_{\alpha}^{a}$ with the fermion fields $\psi_{i}$, $\psi_{j}$,
the scalar fields
$\phi_{k}$, $\phi_{l}$, the pair of gauge bosons $A_{\mu}^{m}$,
$A_{\nu}^{n}$,
the triplet of gauge bosons $A_{\lambda}^{l}$, $A_{\mu}^{m}$,
$A_{\nu}^{n}$
and the scalar and gauge fields $\phi_{k}$, $\phi_{l}$, $A_{\mu}^{m}$
are shown in fig.\ 2.
We work in the class of conventional $R_{\xi}$ gauges.

In the S-matrix formulation of the PT, one considers the S-matrix
elements for the scattering of on-shell fermions.
For the four-fermion process
$\psi_{i}(p)+\psi_{i'}(p')\rightarrow\psi_{j}(p+q)+\psi_{j'}(p'-q)$,
the  S-matrix element $T$ may be decomposed as
\be
T(s,t,m)
=
T_{1}(t) + T_{2}(t,m) + T_{3}(s,t,m)
\ee
where $s = (p+p')^{2}$ and $t = q^{2}$. The component
$T_{1}$, depending (up to trivial external wavefunctions)
only on the momentum transfer $q^{2}$, is the full
propagator-like component of the process.
Because of their different dependences on the kinematic variables
$s$ and $t$ and the fermion mass $m$, the components $T_{1}$,
$T_{2}$ and $T_{3}$ must be individually gauge-independent.

The gauge-dependent contribution
to the S-matrix element of the diagram in
fig.\ 1(a) involving the conventional gauge boson two-point function
$\Pi(\xi,q^{2})$ is given by
\be\label{fig1a}
{\rm Fig.\,\,1(a)}
=
\Bigl(\overline{u}_{j'}ig\gamma^{\mu}T_{j'i'}^{a}u_{i'}\Bigr)
\,\frac{-i}{q^{2}}\,i\Pi(\xi,q^{2})\,\frac{-i}{q^{2}}\,
\Bigl(\overline{u}_{j}ig\gamma_{\mu}T_{ji}^{a}u_{i}\Bigr)
\ee
(the particular indices $i,j,i',j'$ are not summed).
The effect of the PT algorithm is to extract the
contributions (the pinch parts)
of the diagrams in figs.\ 1(b), (d) and (f) which have
exactly the propagator-like
form of eq.\ (\ref{fig1a}), i.e.\ a function of $q^{2}$
between two tree level vector--fermion--fermion vertices.
Adding these pinch contributions to the contribution
eq.\ (\ref{fig1a}) gives the component $T_{1}(q^{2})$ of the S-matrix
element and defines the PT ``effective'' two-point function
$\hat{\Pi}(q^{2})$:
\be
T_{1}(q^{2})
=
\Bigl(\overline{u}_{j'}ig\gamma^{\mu}T_{j'i'}^{a}u_{i'}\Bigr)
\,\frac{-i}{q^{2}}\,i\hat{\Pi}(q^{2})\,\frac{-i}{q^{2}}\,
\Bigl(\overline{u}_{j}ig\gamma_{\mu}T_{ji}^{a}u_{i}\Bigr).
\ee
Because the component $T_{1}$ is gauge-independent, so must be
the function $\hat{\Pi}(q^{2})$. The PT gauge-independent one-loop
``effective''
gauge boson self-energy tensor is then given by
\be
\hat{\Pi}_{\mu\nu}^{ab}(q)
=
\biggl(g_{\mu\nu} - \frac{q_{\mu}q_{\nu}}{q^{2}}\biggr)
\delta^{ab}\hat{\Pi}(q^{2}).
\ee

The gauge boson on, say, the r.h.s.\ of fig.\ 1(a), rather than
coupling to
the fermion pair $\psi_{i}$, $\psi_{j}$, may however couple instead
at tree level to a pair of scalar fields
$\phi_{k}$, $\phi_{l}$, or a pair of gauge bosons
$A_{\mu}^{m}$, $A_{\nu}^{n}$,
or a triplet of gauge bosons
$A_{\lambda}^{l}$, $A_{\mu}^{m}$, $A_{\nu}^{n}$,
or a triplet of scalar and gauge fields $\phi_{k}$, $\phi_{l}$,
$A_{\mu}^{m}$.
Thus, at the one-loop level, rather than considering the process
$\psi_{i}\psi_{i'} \rightarrow \psi_{j}\psi_{j'}$,
we can instead consider the process
$\phi_{k}\psi_{i'} \rightarrow \phi_{l}\psi_{j'}$, or
$A_{\mu}^{m}\psi_{i'} \rightarrow A_{\nu}^{n}\psi_{j'}$, or
$A_{\lambda}^{l}\psi_{i'} \rightarrow A_{\mu}^{m}A_{\nu}^{n}\psi_{j'}$,
or $\phi_{k}\psi_{i'} \rightarrow \phi_{l}A_{\mu}^{m}\psi_{j'}$.
In each case, the
contribution to the corresponding S-matrix element
of the diagram analogous to fig.\ 1(a) involving
the conventional gauge boson two-point function is given by
eq.\ (\ref{fig1a}) with the vector--fermion--fermion
vertex term $\overline{u}_{j}ig\gamma_{\mu}T_{ji}^{a}u_{i}$
replaced by the corresponding Feynman rule from fig.\ 2, together with
any appropriate polarization vectors. Clearly, the gauge dependence of
this diagram is independent of the
particular on-shell external particles.

For the given set of external particles, there
then remain the one-loop diagrams
analogous to the vertex and box corrections figs.\ 1(b), (d) and (f) in
the four-fermion case.
Just as in the four-fermion case, these diagrams
implicitly contain propagator-like components (pinch parts), defined
as the components proportional to functions of $q^{2}$ between
the appropriate tree level Feynman rules of fig.\ 2.
These pinch parts may be added to the contribution
analogous to eq.\ (\ref{fig1a}) from the
conventional two-point function to specify the ``effective'' two-point
function for the process, giving the full propagator-like
contribution to the corresponding S-matrix element.
{\em The requirement of universality of the PT gauge boson self-energy
$\hat{\Pi}_{\mu\nu}^{ab}(q)$ is that it be this ``effective''
two-point function, independent of the species of external particle.}
We will now show that this is indeed the case.

\vspace{10pt}

{\bf 3.} In the S-matrix PT with external fermions,
the identification of the pinch parts is made using
the elementary Ward identity
\bea
\ks
&=&
(\ps - m) - (\ps - \ks - m) \\
\label{fermionid}
&=& S^{-1}(p;m) - S^{-1}(p-k;m)
\eea
in the numerator of the Feynman integral for a given diagram,
where $k_{\mu}$ is the four-momentum carried away by the gauge boson
and $p_{\mu}$ and $(p - k)_{\mu}$ are the
four-momenta of the adjacent fermions.
In general, in a conventional $R_{\xi}$ gauge, such factors
of four-momentum $k_{\mu}$ arise both from the longitudinal components
of the gauge field propagators and from triple gauge vertices.
For example, in fig.\ 1(b), with $p_{\mu}$ the four-momentum of the
incoming fermion $\psi_{i}$, the fermion line inside the loop
has propagator $iS(p-k;m)$.
Using the Ward identity eq.\ (\ref{fermionid}) for a four-momentum
factor $k_{\mu}$ contracted with the Dirac matrix
$\gamma^{\mu}$ associated with the lower vector--fermion--fermion
vertex, the first term from eq.\ (\ref{fermionid})
gives zero contribution since $(\ps - m)u_{i}(p) = 0$
for the on-shell fermion, while the
second term exactly cancels the fermion propagator, so giving
a pinch part. A similar effect occurs
in the upper vertex for a four-momentum
factor $(q+k)_{\nu}$ contracted with $\gamma^{\nu}$
and acting on $\overline{u}_{j}(p+q)$.
Thus, by using the Ward identity eq.\ (\ref{fermionid}),
the total propagator-like contribution of the diagram may be found.
The propagator-like contributions of the
diagrams shown in figs.\ 1(d) and (f) are
identified in a similar way,
with in these cases four-momentum
factors coming only from the gauge propagators.

Once it is established that the quantity being calculated is
gauge-independent,
one is at liberty to choose the gauge in which the
calculation is simplest. Choosing the Feynman gauge $\xi = 1$,
the gauge field propagators $iD_{\mu\nu}^{ab}$
are proportional to $g_{\mu\nu}$ and so
the only possible source of four-momentum factors with which to
generate pinch parts is the triple gauge vertex\footnote{
In the class of background field $R_{\xi}$ gauges, the effect of
choosing the Feynman gauge $\xi_{q} = 1$ for the quantum gauge fields
is to eliminate even the background--quantum--quantum field triple
gauge vertex as a source of such four-momentum factors. This is the
reason behind the recent observations \cite{edrnjw}--\cite{denner2}
that at one loop the background gauge boson two-point function
calculated in the Feynman quantum gauge is identical to the PT
``effective'' two-point function, both in QCD and the SM electroweak
sector. For $\xi_{q} \neq 1$ however, this result no longer holds
since the background gauge boson two-point function is
$\xi_{q}$-dependent. Thus, only for $\xi_{q} = 1$ does the background
field two-point function give the ``effective'' two-point function
between two tree level background gauge boson vertices. Clearly, for
$\xi_{q} \neq 1$, the PT algorithm can be applied in the background
field formulation to construct again this ``effective'' two-point
function \cite{papav7}.}.
Thus, in the
Feynman gauge, the diagrams in figs.\ 1(d) and (f) have vanishing
pinch parts\footnote{For the case of massless gauge bosons,
the pinch part fig.\ 1(e) of the diagram
fig.\ 1(d) in fact vanishes for all values of $\xi$
in dimensional regularization.} and the entire pinch contribution to the
PT self-energy is given by the pinch part of the diagram in fig.\ 1(b).

\pagebreak

For the cases of external scalars and gauge bosons, pinch terms are
generated by similar factors of four-momentum. For the scalars, the
pinch process is exactly similar to the fermion case.
However, in the case of
gauge bosons, there are also pinch terms which arise from the additional
triple gauge vertices involved. These additional pinch terms
are the novel feature of the application of the PT to the case of
external gauge bosons, and must be taken into account.

The demonstration of the universality of the PT gauge boson
self-energy consists in choosing the Feynman gauge and then showing
that, for each of the possible sets of external fields
$\psi_{i}\psi_{j}$, $\phi_{k}\phi_{l}$,
$A_{\mu}^{m}A_{\nu}^{n}$, $A_{\lambda}^{l}A_{\mu}^{m}A_{\nu}^{n}$ and
$\phi_{k}\phi_{l}A_{\mu}^{m}$ to
which the gauge boson $A_{\alpha}^{a}$ couples at tree level,
the pinch part of the diagram analogous to fig.\ 1(b) is the same.
The relevant half of
these diagrams and their pinch part\footnote{
It is emphasized that the vertices
involving four gauge bosons in figs.\ 3(d) and (f)
are {\em not} the usual tree level vertices from the lagrangian
eq.\ (\ref{Lcl}).}
are shown in fig.\ 3. It is easy to show that all
other one-loop diagrams\footnote{The one-loop diagrams involving the
tree level $\phi\phi\phi\phi$, $AAAA$ or
$\phi\phi AA$ vertices with a pair of fields from the four-boson
vertex appearing as external fields result in contributions which do
not have the kinematic form of a function of $q^{2}$ between two of
the appropriate tree level vertices of fig.\ 2.}
have zero pinch part in this gauge, so that the universality
is then proved.

In each case, the gauge boson $A_{\beta}^{b}$ ($A_{\gamma}^{c}$)
in the lower (upper) part of the loop is taken to have
four-momentum $k$ ($-k\!-\!q$) flowing into the triple gauge vertex.
This vertex may be decomposed as originally suggested by 't Hooft
\cite{thooft}:
\be
\Gamma_{\alpha\beta\gamma}^{abc}
=
f^{abc}\Bigl(
\Gamma_{\alpha\beta\gamma}^{F} + \Gamma_{\alpha\beta\gamma}^{P}
\Bigr)
\ee
where
\bea
\Gamma_{\alpha\beta\gamma}^{F}(q,k,-q-k)
&=&
(2k+q)_{\alpha}g_{\beta\gamma}
- 2q_{\beta}g_{\gamma\alpha} + 2q_{\gamma}g_{\alpha\beta}  \\
\Gamma_{\alpha\beta\gamma}^{P}(q,k,-q-k)
&=&
- k_{\beta}g_{\gamma\alpha} - (k+q)_{\gamma}g_{\alpha\beta}.
\eea
The part $\Gamma_{\alpha\beta\gamma}^{F}$
gives no pinch contribution
and obeys a simple QED-like Ward identity
$q^{\alpha}\Gamma_{\alpha\beta\gamma}^{F}(q,k,-q-k)
= [k^{2} - (q+k)^{2}]g_{\beta\gamma}$
involving the difference of two inverse gauge field propagators in the
Feynman gauge. The part $\Gamma_{\alpha\beta\gamma}^{P}$
gives two pinch contributions, one from $k_{\beta}$, the other from
$(k+q)_{\gamma}$.

\vspace{10pt}

i) External fermion pair $\psi_{i}$, $\psi_{j}$

\noindent
For the case of a pair of external fermion fields
$\psi_{i}$, $\psi_{j}$,
the subamplitude for the r.h.s.\ of the diagram fig.\ 3(a) is
\be\label{3arhs}
ig\gamma_{\gamma}T_{jr}^{c}
\frac{i}{\ps - \ks - m + i\epsilon}
ig\gamma_{\beta}T_{ri}^{b}
\ee
(for brevity we omit the spinors).
The pinch parts of this subamplitude are generated by
factors of four-momentum $k^{\beta}$ and $(k + q)^{\gamma}$
multiplying this expression. Using
$\ks = \ps - m - (\ps - \ks - m)$
and $(\ps -m)u_{i}(p) = 0$ for the on-shell incoming
fermion, the pinch part due to a factor $k^{\beta}$ is
\be
ig^{2}\gamma_{\gamma}(T^{c}T^{b})_{ji}.
\ee
Similarly, using $\ks + \qs = \ps + \qs - m - (\ps - \ks - m)$
and $\overline{u}_{j}(p+q)(\ps + \qs - m) = 0$ for the on-shell outgoing
fermion, the pinch part due to a factor $(k+q)^{\gamma}$ is
\be
ig^{2}\gamma_{\beta}(T^{c}T^{b})_{ji}.
\ee

Adding these contributions due to the part
$\Gamma_{\alpha\beta\gamma}^{P}$
of the triple gauge vertex
and using $f^{abc}T^{c}T^{b} = -\frac{1}{2}iN T^{a}$
gives the pinch part fig.\ 3(b) of the diagram fig.\ 3(a):
\be\label{3apinch}
{\rm Fig.\,\,3(a)}|_{\rm pinch}
=
-ig^{2}N\mu^{2\epsilon}\int\frac{d^{n}k}{(2\pi)^{n}}
\frac{1}{k^{2}(k+q)^{2}}
\,\,ig\gamma_{\alpha}T_{ji}^{a}
\ee
(we use always dimensional regularization in
$n = 4-2\epsilon$ dimensions and with 't Hooft mass scale $\mu$).

\vspace{10pt}

ii) External scalar pair $\phi_{k}$, $\phi_{l}$

\noindent
For the case of a pair of external scalar fields
$\phi_{k}$, $\phi_{l}$,
the subamplitude for the r.h.s.\ of the diagram fig.\ 3(c) is
\be\label{3crhs}
ig(2p-k+q)_{\gamma}R_{lr}^{c}
\frac{i}{(p-k)^{2} - M^{2} + i\epsilon}
ig(2p-k)_{\beta}R_{rk}^{b}.
\ee
As in the case of fermions, the pinch parts of this subamplitude are
generated by factors of four-momentum $k^{\beta}$ and $(k + q)^{\gamma}$
multiplying this expresion. Using
$k^{\beta}(2p-k)_{\beta} = p^{2} - M^{2} -[(p-k)^{2} - M^{2}]$
and $p^{2} - M^{2} = 0$ for the on-shell incoming scalar particle,
the pinch part due to a factor $k^{\beta}$ is
\be
ig^{2}(2p-k+q)_{\gamma}(R^{c}R^{b})_{lk}.
\ee
Similarly, using
$(k+q)^{\gamma}(2p-k+q)_{\gamma} = (p+q)^{2} - M^{2}
-[(p-k)^{2} - M^{2}]$
and $(p+q)^{2} - M^{2} = 0$ for the on-shell outgoing scalar particle,
the pinch part due to a factor $(k+q)^{\gamma}$ is
\be
ig^{2}(2p-k)_{\beta}(R^{c}R^{b})_{lk}.
\ee

Adding these contributions due to the part
$\Gamma_{\alpha\beta\gamma}^{P}$
of the triple gauge vertex
and using $f^{abc}R^{c}R^{b} = -\frac{1}{2}iN R^{a}$
gives the pinch part fig.\ 3(d) of the diagram fig.\ 3(c):
\be\label{3cpinch}
{\rm Fig.\,\,3(c)}|_{\rm pinch}
=
-ig^{2}N\mu^{2\epsilon}\int\frac{d^{n}k}{(2\pi)^{n}}
\frac{1}{k^{2}(k+q)^{2}}
\,\,ig(2p+q)_{\alpha}R_{lk}^{a}.
\ee

\vspace{10pt}

\pagebreak

iii) External gauge boson pair $A_{\mu}^{m}$, $A_{\nu}^{n}$

\noindent
For the case of a pair of external gauge boson fields
$A_{\mu}^{m}$, $A_{\nu}^{n}$,
the subamplitude for the r.h.s.\ of the diagram fig.\ 3(e) is
\bea
& &
gf^{ncr}[
  (2k-p+q)_{\nu}g_{\gamma}^{\phantom{\gamma}\rho}
+ (2p-k+q)_{\gamma}g_{\phantom{\rho}\nu}^{\rho}
+ (-p-k-2q)^{\rho}g_{\nu\gamma}
] \nonumber\\
\label{3erhs}
& &
\times\frac{-i}{(p-k)^{2}}gf^{mrb}[
  (2k-p)_{\mu}g_{\rho\beta}
+ (-p-k)_{\rho}g_{\beta\mu}
+ (2p-k)_{\beta}g_{\mu\rho} ].
\eea
Exactly as in the fermion and scalar cases, pinch contributions are
generated by factors of $k^{\beta}$ and $(q+k)^{\gamma}$ multipying
this expression. Using $p^{2} = 0$ and $(p+q)^{2} = 0$ for the
on-shell incoming and outgoing external gauge bosons,
these factors give respectively
\be
ig^{2}f^{ncr}f^{mrb}[
  (2k-p+q)_{\nu}g_{\gamma}^{\phantom{\gamma}\rho}
+ (2p-k+q)_{\gamma}g_{\phantom{\rho}\nu}^{\rho}
+ (-p-k-2q)^{\rho}g_{\nu\gamma}
]g_{\mu\rho}
\ee
and
\be
ig^{2}f^{ncr}f^{mrb}[
  (2k-p)_{\mu}g_{\rho\beta}
+ (-p-k)_{\rho}g_{\beta\mu}
+ (2p-k)_{\beta}g_{\mu\rho}
]g_{\phantom{\rho}\nu}^{\rho}.
\ee
However, there is also a pinch term generated within the subamplitude
itself from the contraction
$(-p-k-2q)^{\rho}(-p-k)_{\rho} = (p-k)^{2} + 4kp + 2pq + 2qk$
in eq.\ (\ref{3erhs}). This generates a pinch term
\be
-ig^{2}f^{ncr}f^{mrb}g_{\nu\gamma}g_{\beta\mu}
\ee
independent of any other factors in the overall amplitude.

Adding the two contributions due to the part
$\Gamma_{\alpha\beta\gamma}^{P}$
of the triple gauge vertex
and the contribution proportional
to the full triple gauge vertex
$\Gamma_{\alpha\beta\gamma}$
and using the identity
$f^{abc}f^{ncr}f^{mrb} = \frac{1}{2}Nf^{anm}$
gives the pinch part fig.\ 3(f) of the diagram fig.\ 3(e):
\bea
{\rm Fig.\,\,3(e)}|_{\rm pinch}
&=&
-ig^{2}N\mu^{2\epsilon}\int\frac{d^{n}k}
{(2\pi)^{n}}\frac{1}{k^{2}(k+q)^{2}} \nonumber \\
\label{3epinch}
& &
\times gf^{amn}[
  (2p+q)_{\alpha}g_{\mu\nu}
+ (-2q-p)_{\mu}g_{\nu\alpha}
+ (q-p)_{\nu}g_{\alpha\mu} ].
\eea

\vspace{10pt}

iv) External gauge boson triple $A_{\lambda}^{l}$,
$A_{\mu}^{m}$, $A_{\nu}^{n}$

\noindent
For the case of three external gauge boson fields
$A_{\lambda}^{l}$, $A_{\mu}^{m}$, $A_{\nu}^{n}$, there are three
distinct diagrams figs.\ 3(g), (h) and (i) which
must be taken into account.
For the first of these, fig.\ 3(g), the
subamplitude for the r.h.s.\ of the diagram is
\bea
& &
\hspace*{-20pt}
-ig^{2}[
  f^{rln}f^{rcd}( g_{\lambda\gamma}g_{\nu\delta}
                 -g_{\lambda\delta}g_{\nu\gamma} )
+ f^{rld}f^{rnc}( g_{\lambda\nu}g_{\gamma\delta}
                 -g_{\lambda\gamma}g_{\nu\delta} )
+ f^{rlc}f^{rnd}( g_{\lambda\nu}g_{\gamma\delta}
                 -g_{\lambda\delta}g_{\nu\gamma} ) ] \nonumber\\
& &
\hspace*{-20pt}
\times \frac{-i}{(p-k)^{2}}
gf^{bmd}[ (2k-p)_{\mu}g_{\phantom{\delta}\beta}^{\delta}
         +(2p-k)_{\beta}g_{\mu}^{\phantom{\mu}\delta}
         +(-p-k)^{\delta}g_{\beta\mu} ].
\eea
The pinch part of this subamplitude is generated by a factor
$k^{\beta}$ from $\Gamma_{\alpha\beta\gamma}^{P}$, which,
using\hfill

\pagebreak

\noindent
$p^{2} = 0$ for the on-shell gauge boson $A_{\mu}^{m}$, gives
\bea
g^{4}f^{abc}f^{bmd}[
  f^{rln}f^{rcd}( g_{\lambda\alpha}g_{\nu\mu}
                 -g_{\lambda\mu}g_{\nu\alpha} )\,\,\,\,\,\,\,\,\,
& & \nonumber \\
+ f^{rld}f^{rnc}( g_{\lambda\nu}g_{\alpha\mu}
                 -g_{\lambda\alpha}g_{\nu\mu} )\,\,\,\,\,
& & \nonumber \\
\label{3grhspinch}
+ f^{rlc}f^{rnd}( g_{\lambda\nu}g_{\alpha\mu}
                 -g_{\lambda\mu}g_{\nu\alpha} ) ].
\eea
For the second of the diagrams, fig.\ 3(h), the subamplitude
for the r.h.s.\ of the diagram is
\bea
& &
\hspace*{-21pt}
-ig^{2}[
  f^{rmb}f^{rld}( g_{\mu\lambda}g_{\beta\delta}
                 -g_{\mu\delta}g_{\beta\lambda} )
+ f^{rmd}f^{rbl}( g_{\mu\beta}g_{\lambda\delta}
                 -g_{\mu\lambda}g_{\beta\delta} )
+ f^{rml}f^{rbd}( g_{\mu\beta}g_{\lambda\delta}
                 -g_{\mu\delta}g_{\beta\lambda} ) ] \nonumber \\
& &
\hspace*{-20pt}
\times\frac{-i}{(p - k + r)^{2}}
gf^{cdn}[ (-p\!-\!2q\!-\!r\!-\!k)^{\delta}g_{\nu\gamma}
+(2p\!-\!k\!+\!q\!+\!2r)_{\gamma}g_{\phantom{\delta}\nu}^{\delta}
+(2k\!-\!p\!+\!q\!-\!r )_{\nu  }g_{\gamma}^{\phantom{\gamma}\delta} ].
\eea
The pinch part of this subamplitude is generated by a factor
$(k+q)^{\gamma}$ from $\Gamma_{\alpha\beta\gamma}^{P}$, which, using
$(p+q+r)^{2} = 0$ for the on-shell gauge boson $A_{\nu}^{n}$, gives
\bea
g^{4}f^{abc}f^{cdn}[
  f^{rmb}f^{rld}( g_{\mu\lambda}g_{\alpha\nu}
                 -g_{\mu\nu}g_{\alpha\lambda} )\,\,\,\,\,\,\,\,\,
& & \nonumber \\
+ f^{rmd}f^{rbl}( g_{\mu\alpha}g_{\lambda\nu}
                 -g_{\mu\lambda}g_{\alpha\nu} )\,\,\,\,\,
& & \nonumber \\
\label{3hrhspinch}
+ f^{rml}f^{rbd}( g_{\mu\alpha}g_{\lambda\nu}
                 -g_{\mu\nu}g_{\alpha\lambda} ) ].
\eea
Lastly, for the diagram fig.\ 3(i), the subamplitude for the r.h.s.\ is
\bea
& &
\hspace*{-20pt}
-ig^{2}[
  f^{rmn}f^{rcd}( g_{\mu\gamma}g_{\nu\delta}
             \!-\!g_{\mu\delta}g_{\nu\gamma} )
+ f^{rmd}f^{rnc}( g_{\mu\nu}g_{\gamma\delta}
             \!-\!g_{\mu\gamma}g_{\nu\delta} )
+ f^{rmc}f^{rnd}( g_{\mu\nu}g_{\gamma\delta}
             \!-\!g_{\mu\delta}g_{\nu\gamma} ) ] \nonumber\\
& &
\hspace*{-20pt}
\times \frac{-i}{(r-k)^{2}}
gf^{bld}[ (-k-r)^{\delta}g_{\beta\lambda}
         +(2r-k)_{\beta}g_{\lambda}^{\phantom{\lambda}\delta}
         +(2k-r)_{\lambda}g_{\phantom{\delta}\beta}^{\delta} ].
\eea
The pinch term for this subamplitude is generated by a factor
$k^{\beta}$ from $\Gamma_{\alpha\beta\gamma}^{P}$, which, using
$r^{2} = 0$ for the on-shell gauge boson $A_{\lambda}^{l}$, gives
\bea
g^{4}f^{abc}f^{bld}[
  f^{rmn}f^{rcd}( g_{\mu\alpha}g_{\nu\lambda}
                 -g_{\mu\lambda}g_{\nu\alpha} )\,\,\,\,\,\,\,\,\,
& & \nonumber \\
+ f^{rmd}f^{rnc}( g_{\mu\nu}g_{\alpha\lambda}
                 -g_{\mu\alpha}g_{\nu\lambda} )\,\,\,\,\,
& & \nonumber \\
\label{3irhspinch}
+ f^{rmc}f^{rnd}( g_{\mu\nu}g_{\alpha\lambda}
                 -g_{\mu\lambda}g_{\nu\alpha} ) ].
\eea
Defining the group-theoretic quantity
\be
f(abcd)
=
f^{akl}f^{blm}f^{cmn}f^{dnk}
\ee
there then exist the following identities:
\bea\label{fid2}
f(abcd) = f(bcda)
&=&
f(badc) \\
\label{fid3}
f(abcd) - f(abdc)
&=&
-\textstyle{\frac{1}{2}}Nf^{abn}f^{cdn}.
\eea
Using these identities, the three pinch terms proportional to
eqs.\ (\ref{3grhspinch}), (\ref{3hrhspinch}) and
(\ref{3irhspinch}) may be
combined to give the pinch part fig.\ 3(j) of the diagrams figs.\ 3(g),
(h) and (i):
\bea
{\rm Figs.\,\,3(g)+(h)+(i)}|_{\rm pinch}
&=&
-ig^{2}N\mu^{2\epsilon}\int\frac{d^{n}k}
{(2\pi)^{n}}\frac{1}{k^{2}(k+q)^{2}} \nonumber \\
& &\label{3ghipinch}
\times-ig^{2}[
  f^{ral}f^{rmn}( g_{\mu\alpha}g_{\nu\lambda}
                 -g_{\mu\lambda}g_{\nu\alpha} ) \nonumber \\
& &
\hspace*{35pt}
+ f^{ran}f^{rlm}( g_{\lambda\alpha}g_{\mu\nu}
                 -g_{\lambda\nu}g_{\mu\alpha} ) \nonumber \\
& &
\hspace*{35pt}
+ f^{ram}f^{rln}( g_{\lambda\alpha}g_{\mu\nu}
                 -g_{\mu\lambda}g_{\alpha\nu} ) ].
\eea

\vspace{10pt}

v) External scalar and gauge boson triple $\phi_{k}$, $\phi_{l}$,
$A_{\mu}^{m}$

\noindent
Finally, for the case of three external scalar and gauge boson fields
$\phi_{k}$, $\phi_{l}$, $A_{\mu}^{m}$, there are again three diagrams,
figs.\ 3(k), (l) and (m), which contribute. In an exactly similar way
to the previous cases, pinch parts are generated by factors
$k^{\beta}$ and $(k+q)^{\gamma}$ from $\Gamma_{\alpha\beta\gamma}^{P}$.
Adding these contributions gives the pinch part fig.\ 3(n) of the
diagrams figs.\ 3(k), (l) and (m):
\bea
{\rm Figs.\,\,3(k)+(l)+(m)}|_{\rm pinch}
&=&
-ig^{2}N\mu^{2\epsilon}\int\frac{d^{n}k}
{(2\pi)^{n}}\frac{1}{k^{2}(k+q)^{2}} \nonumber \\
& &\label{3klmpinch}
\times -ig^{2}\{R^{a},R^{m}\}_{lk}\,g_{\alpha\mu}.
\eea

\vspace{10pt}

We see that in each case, the pinch part
eqs.\ (\ref{3apinch}), (\ref{3cpinch}), (\ref{3epinch}),
(\ref{3ghipinch}) and (\ref{3klmpinch})
is given by the tree level Feynman rule from fig.\ 2
for the gauge boson $A_{\alpha}^{a}$ coupling to the given external
particles, multiplied by the same
function of $q^{2}$. Given that all other diagrams
have zero pinch part, the universality is therefore proved.

The gauge-independent universal PT self-energy is now obtained by
adding the pinch
contribution, multiplied by an inverse propagator
$iq^{2}$ and a factor two (since it occurs
on each side of the diagram), to the conventional self-energy:
\be
i\hat{\Pi}(q^{2})
=
i\Pi(\xi = 1,q^{2})
+
2Ng^{2}
\mu^{2\epsilon}\int\frac{d^{n}k}{(2\pi)^{n}}\frac{q^{2}}{k^{2}(k+q)^{2}}.
\ee
At asymptotic $q^{2}$, this PT self-energy has the behaviour expected
from the RG $\beta$ function.

\vspace{10pt}

{\bf 4.} It has been shown here how the S-matrix PT may be extended to
include as external particles all five of the different combinations
$\psi_{i}\psi_{j}$, $\phi_{k}\phi_{l}$, $A_{\mu}^{m}A_{\nu}^{n}$,
$A_{\lambda}^{l}A_{\mu}^{m}A_{\nu}^{n}$ and
$\phi_{k}\phi_{l}A_{\mu}^{m}$ to which the gauge boson
$A_{\alpha}^{a}$ couples at tree level. The universality of the PT
one-loop ``effective'' gauge boson two-point function has then been
demonstrated explicitly.

The interest of this result is that it indicates how the concept of a
gauge-independent effective charge $g(q^{2})$ valid at all
$q^{2}$, not just the
asymptotic region governed by the RG $\beta$ function, may be extended
from abelian QED, where it arises naturally, to non-abelian gauge
theories. The crucial point is the recognition that the required
quantity is the ``effective'' two-point function
$\hat{\Pi}_{\mu\nu}^{ab}$ defined between
any two tree level {\em vertices} involving $A_{\mu}^{a}$ and
$A_{\nu}^{b}$.
That this is precisly the quantity constructed in the PT is the PT's
distinguishing feature. In order, however, to promote this effective
charge to a generally applicable
method for accounting for a  well-defined,
infinite subset of gauge-independent diagrams,
it is necessary to demonstrate that the PT ``effective'' two-point
function remains simulaneously gauge-independent and universal
in processes involving {\em off-shell}
external fields. If, as seems likely, both these properties persist,
then this effective charge would have immediate applications in QCD
renormalon calculus and, when extended to broken theories,
in electroweak phenomenology. Work is under way in this direction.

\vspace{10pt}

I wish to thank Eduardo de Rafael for many useful discussions.
This work was supported by EC grant ERB4001GT933989.

\pagebreak

{\bf Figure Captions}

\vspace{10pt}

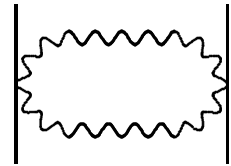
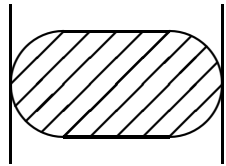
{\bf Fig.\ 1}. (a) The conventional one-loop gauge boson two-point
function contribution to the four-fermion process
$\psi_{i}\psi_{i'}\rightarrow\psi_{j}\psi_{j'}$. (b)--(g) The remaining
one-loop diagrams involving gauge bosons (the external leg corrections
associated with (d) are not shown), together with their pinch parts.
(h) The PT gauge-independent ``effective'' gauge boson two-point
function.

\vspace{10pt}

{\bf Fig.\ 2}. The Feynman rules for the five different sets of fields
to which a single gauge boson $A_{\alpha}^{a}$ couples at tree level.

\vspace{10pt}

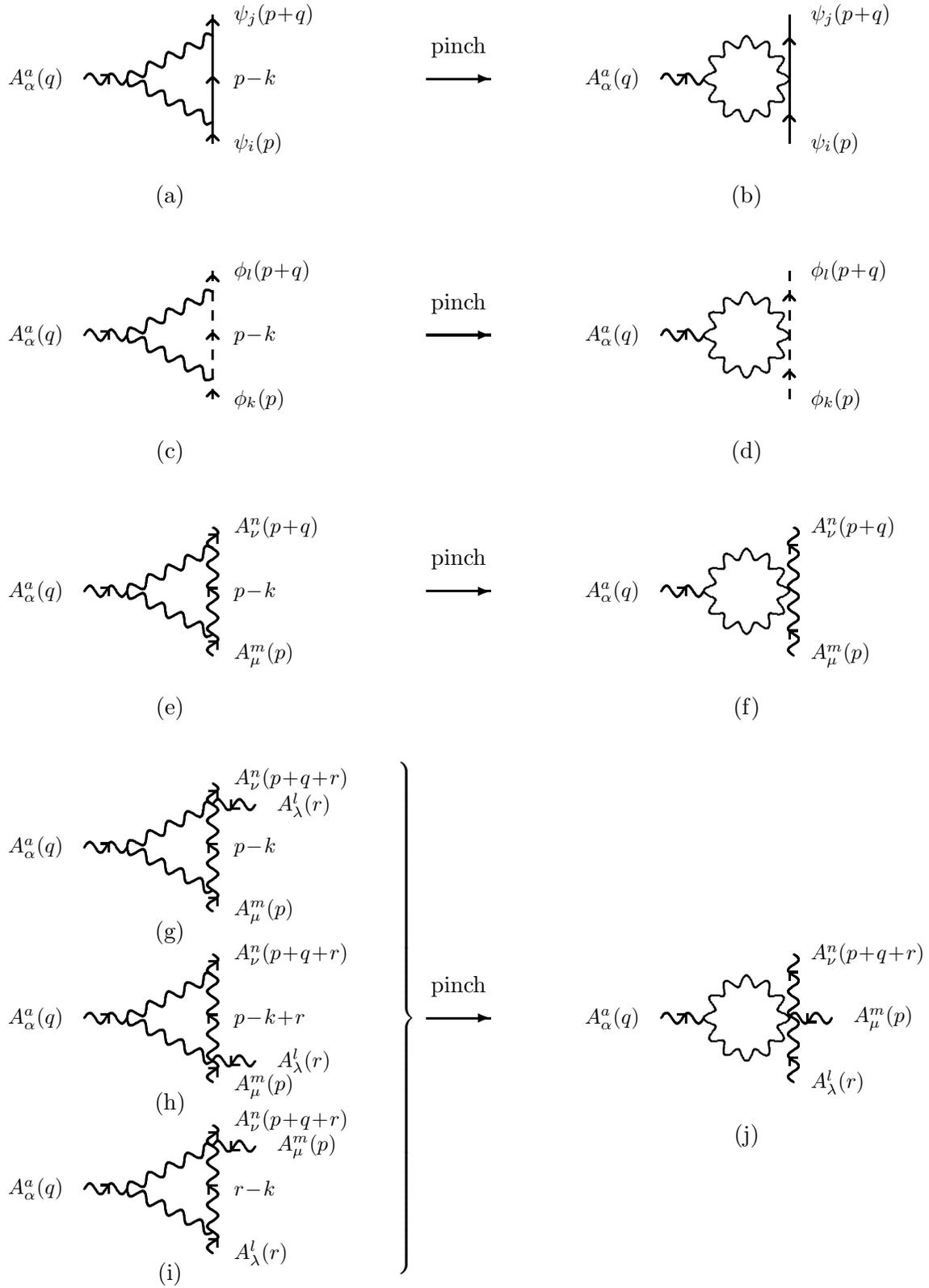
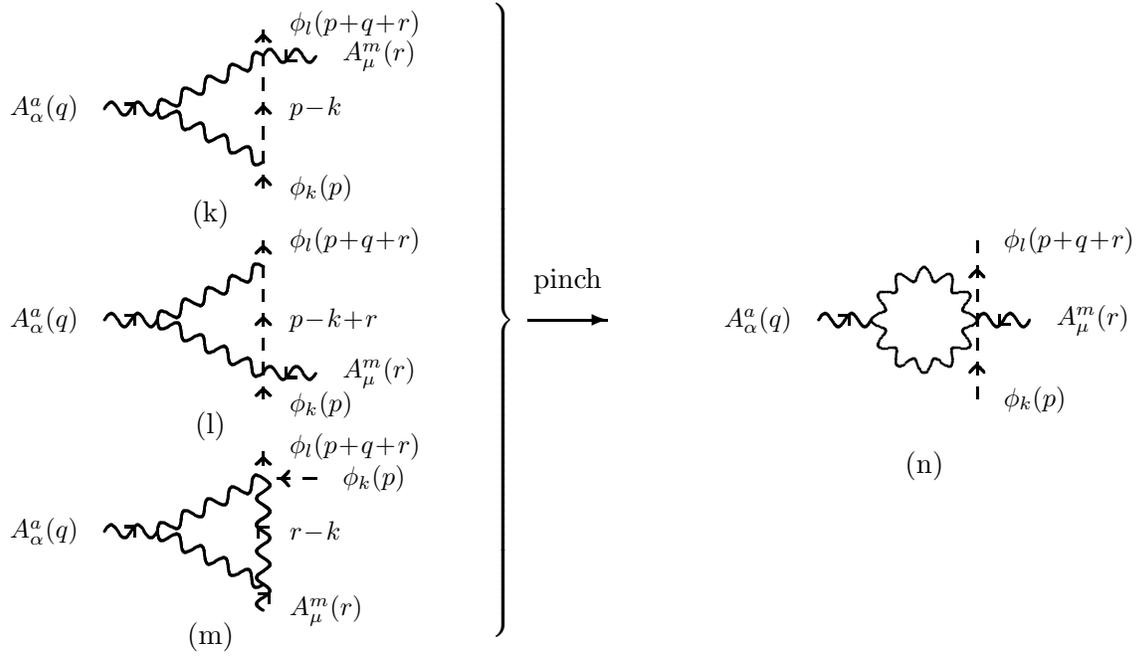
{\bf Fig.\ 3}. The Feynman diagrams giving pinch parts in the
Feynman gauge for the five different sets of external fields
$\psi_{i}\psi_{j}$, $\phi_{k}\phi_{l}$, $A_{\mu}^{m}A_{\nu}^{n}$,
$A_{\lambda}^{l}A_{\mu}^{m}A_{\nu}^{n}$ and
$\phi_{k}\phi_{l}A_{\mu}^{m}$.

\pagebreak

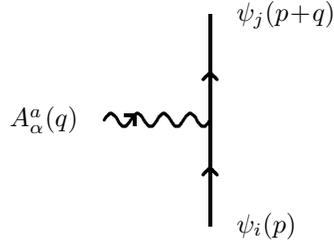
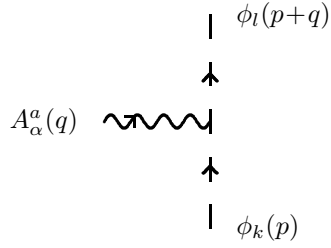
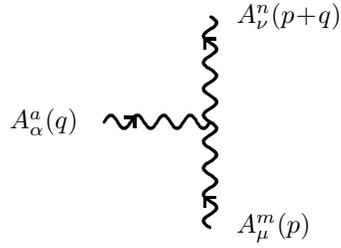
\noindent
\begin{picture}(425,595)(73,0)

\thicklines


\put(235,560){\makebox(0,0)[r]{$\psi_{j'}(p'\!-\!q)$}}
\put(235,500){\makebox(0,0)[r]{$\psi_{i'}(p')$}}
\put(245,500){\line(0,1){60}}
\multiput(245,515)(-0.1,-0.1){28}{\circle*{0.2}}
\multiput(245,515)( 0.1,-0.1){28}{\circle*{0.2}}
\multiput(245,548)(-0.1,-0.1){28}{\circle*{0.2}}
\multiput(245,548)( 0.1,-0.1){28}{\circle*{0.2}}
\put(245,530){\usebox{\gluon}}
\put(285,530){\circle{40}}
\put(275,540){\line(1,1){10}}
\put(275,540){\line(-1,-1){10}}
\put(280,535){\line(1,1){13.2}}
\put(280,535){\line(-1,-1){13.2}}
\put(285,530){\line(1,1){14.1}}
\put(285,530){\line(-1,-1){14.1}}
\put(290,525){\line(1,1){13.2}}
\put(290,525){\line(-1,-1){13.2}}
\put(295,520){\line(1,1){10}}
\put(295,520){\line(-1,-1){10}}
\put(305,530){\usebox{\gluon}}
\put(325,500){\line(0,1){60}}
\multiput(325,515)(-0.1,-0.1){28}{\circle*{0.2}}
\multiput(325,515)( 0.1,-0.1){28}{\circle*{0.2}}
\multiput(325,548)(-0.1,-0.1){28}{\circle*{0.2}}
\multiput(325,548)( 0.1,-0.1){28}{\circle*{0.2}}
\put(285,480){\makebox(0,0)[c]{(a)}}
\put(335,560){\makebox(0,0)[l]{$\psi_{j}(p\!+\!q)$}}
\put(335,500){\makebox(0,0)[l]{$\psi_{i}(p)$}}


\put(120,390){\line(0,1){60}}
\put(120,420){\usebox{\gluon}}
\put(140,420){\usebox{\gluon}}
\put(160,420){\usebox{\trgluon}}
\put(160,420){\usebox{\brgluon}}
\put(200,390){\line(0,1){60}}
\put(160,375){\makebox(0,0)[c]{\footnotesize +\,\,reversed\,\,diagram}}
\put(160,360){\makebox(0,0)[c]{(b)}}

\put(285,435){\makebox(0,0)[c]{pinch}}
\put(270,420){\vector(1, 0){30}}


\put(370,390){\line(0,1){60}}
\put(370,420){\usebox{\gluon}}
\put(390,420){\usebox{\gluon}}
\put(430,420){\usebox{\leftloop}}
\put(430,420){\usebox{\rightloop}}
\put(450,390){\line(0,1){60}}
\put(410,375){\makebox(0,0)[c]{\footnotesize +\,\,reversed\,\,diagram}}
\put(410,360){\makebox(0,0)[c]{(c)}}


\put(120,280){\line(0,1){60}}
\put(120,310){\usebox{\gluon}}
\put(140,310){\usebox{\gluon}}
\put(160,310){\usebox{\gluon}}
\put(180,310){\usebox{\gluon}}
\put(200,280){\line(0,1){60}}
\put(200,310){\usebox{\rightloop}}
\put(160,265){\makebox(0,0)[c]{\footnotesize +\,\,reversed\,\,diagram}}
\put(160,250){\makebox(0,0)[c]{(d)}}

\put(285,325){\makebox(0,0)[c]{pinch}}
\put(270,310){\vector(1, 0){30}}


\put(370,280){\line(0,1){60}}
\put(370,310){\usebox{\gluon}}
\put(390,310){\usebox{\gluon}}
\put(410,310){\usebox{\gluon}}
\put(430,310){\usebox{\gluon}}
\put(450,280){\line(0,1){60}}
\put(470,310){\usebox{\leftloop}}
\put(470,310){\usebox{\rightloop}}
\put(410,265){\makebox(0,0)[c]{\footnotesize +\,\,reversed\,\,diagram}}
\put(410,250){\makebox(0,0)[c]{(e)}}


\put(120,170){\line(0,1){60}}
\put(120,182.5){\usebox{\gluon}}
\put(140,182.5){\usebox{\gluon}}
\put(160,182.5){\usebox{\gluon}}
\put(180,182.5){\usebox{\gluon}}
\put(120,217.5){\usebox{\gluon}}
\put(140,217.5){\usebox{\gluon}}
\put(160,217.5){\usebox{\gluon}}
\put(180,217.5){\usebox{\gluon}}
\put(200,170){\line(0,1){60}}
\put(160,155){\makebox(0,0)[c]{\footnotesize +\,\,crossed\,\,diagram}}
\put(160,140){\makebox(0,0)[c]{(f)}}

\put(285,215){\makebox(0,0)[c]{pinch}}
\put(270,200){\vector(1, 0){30}}


\put(370,170){\line(0,1){60}}
\put(390,200){\usebox{\leftloop}}
\put(390,182.5){\usebox{\bottomgluon}}
\put(410,182.5){\usebox{\bottomgluon}}
\put(390,217.5){\usebox{\topgluon}}
\put(410,217.5){\usebox{\topgluon}}
\put(430,200){\usebox{\rightloop}}
\put(450,170){\line(0,1){60}}
\put(410,140){\makebox(0,0)[c]{(g)}}


\put(285, 80){\oval(80,40)}
\put(245, 50){\line(0,1){60}}
\put(325, 50){\line(0,1){60}}
\put(265, 60){\line(1,1){40}}
\put(275, 60){\line(1,1){38.2}}
\put(285, 60){\line(1,1){34.1}}
\put(295, 60){\line(1,1){28.2}}
\put(305, 60){\line(1,1){20}}
\put(305,100){\line(-1,-1){40}}
\put(295,100){\line(-1,-1){38.2}}
\put(285,100){\line(-1,-1){34.1}}
\put(275,100){\line(-1,-1){28.2}}
\put(265,100){\line(-1,-1){20}}
\put(285, 30){\makebox(0,0)[c]{(h)}}

\put(500,-20){\makebox(0,0)[br]{\LARGE Fig.\ 1}}

\end{picture}

\noindent
\begin{picture}(425,595)(20,-50)

\thicklines


\put(100,510){\usebox{\gluon}}
\put(100,510){\usebox{\rgluonarrow}}
\put(120,510){\usebox{\gluon}}
\put(140,470){\line(0,1){80}}
\multiput(140,492)(-0.1,-0.1){28}{\circle*{0.2}}
\multiput(140,492)( 0.1,-0.1){28}{\circle*{0.2}}
\multiput(140,528)(-0.1,-0.1){28}{\circle*{0.2}}
\multiput(140,528)( 0.1,-0.1){28}{\circle*{0.2}}
\put( 90,510){\makebox(0,0)[r]{\small $A_{\alpha}^{a}(q)$}}
\put(150,550){\makebox(0,0)[l]{\small $\psi_{j}(p\!+\!q)$}}
\put(150,470){\makebox(0,0)[l]{\small $\psi_{i}(p)$}}
\put(232,450){\makebox(0,0)[c]{(a)}}
\put(340,510){\makebox(0,0)[c]{$ig\gamma_{\alpha}T_{ji}^{a}$}}


\put(100,380){\usebox{\gluon}}
\put(100,380){\usebox{\rgluonarrow}}
\put(120,380){\usebox{\gluon}}
\multiput(140,339.5)(0,18){5}{\line(0,1){9}}
\multiput(140,363)(-0.1,-0.1){28}{\circle*{0.2}}
\multiput(140,363)( 0.1,-0.1){28}{\circle*{0.2}}
\multiput(140,398)(-0.1,-0.1){28}{\circle*{0.2}}
\multiput(140,398)( 0.1,-0.1){28}{\circle*{0.2}}
\put( 90,380){\makebox(0,0)[r]{\small $A_{\alpha}^{a}(q)$}}
\put(150,420){\makebox(0,0)[l]{\small $\phi_{l}(p\!+\!q)$}}
\put(150,340){\makebox(0,0)[l]{\small $\phi_{k}(p)$}}
\put(232,320){\makebox(0,0)[c]{(b)}}
\put(345,380){\makebox(0,0)[c]{$ig(2p+q)_{\alpha}R_{lk}^{a}$}}


\put(100,250){\usebox{\gluon}}
\put(100,250){\usebox{\rgluonarrow}}
\put(120,250){\usebox{\gluon}}
\put(140,210){\usebox{\vgluon}}
\put(140,210){\usebox{\ugluonarrow}}
\put(140,230){\usebox{\vgluon}}
\put(140,250){\usebox{\vgluon}}
\put(140,270){\usebox{\vgluon}}
\put(140,270){\usebox{\ugluonarrow}}
\put( 90,250){\makebox(0,0)[r]{\small $A_{\alpha}^{a}(q)$}}
\put(150,290){\makebox(0,0)[l]{\small $A_{\nu}^{n}(p\!+\!q)$}}
\put(150,210){\makebox(0,0)[l]{\small $A_{\mu}^{m}(p)$}}
\put(232,190){\makebox(0,0)[c]{(c)}}
\put(330,270){\makebox(0,0)[c]{$gf^{amn}[( 2p+q)_{\alpha}g_{\mu\nu}$}}
\put(360,250){\makebox(0,0)[c]{$+(-2q-p)_{\mu}g_{\nu\alpha}$}}
\put(370,230){\makebox(0,0)[c]{$+( q- p)_{\nu}g_{\alpha\mu}]$}}


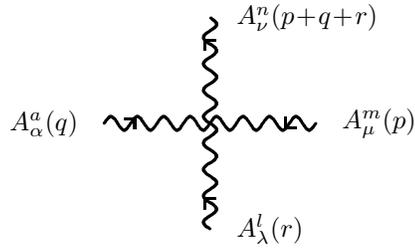
\put(100,120){\usebox{\gluon}}
\put(100,120){\usebox{\rgluonarrow}}
\put(120,120){\usebox{\gluon}}
\put(140,120){\usebox{\gluon}}
\put(160,120){\usebox{\gluon}}
\put(180,120){\usebox{\lgluonarrow}}
\put(140, 80){\usebox{\vgluon}}
\put(140, 80){\usebox{\ugluonarrow}}
\put(140,100){\usebox{\vgluon}}
\put(140,120){\usebox{\vgluon}}
\put(140,140){\usebox{\vgluon}}
\put(140,140){\usebox{\ugluonarrow}}
\put( 90,120){\makebox(0,0)[r]{\small $A_{\alpha}^{a}(q)$}}
\put(150,160){\makebox(0,0)[l]{\small $A_{\nu}^{n}(p\!+\!q\!+\!r)$}}
\put(190,120){\makebox(0,0)[l]{\small $A_{\mu}^{m}(p)$}}
\put(150, 80){\makebox(0,0)[l]{\small $A_{\lambda}^{l}(r)$}}
\put(232, 60){\makebox(0,0)[c]{(d)}}
\put(335,140){\makebox(0,0)[c]{$-ig^{2}[f^{ral}f^{rmn}
( g_{\mu\alpha}g_{\nu\lambda} - g_{\mu\lambda}g_{\nu\alpha} )$}}
\put(360,120){\makebox(0,0)[c]{$+ f^{ran}f^{rlm}
( g_{\lambda\alpha}g_{\mu\nu} - g_{\lambda\nu}g_{\mu\alpha} )$}}
\put(370,100){\makebox(0,0)[c]{$+ f^{ram}f^{rln}
( g_{\lambda\alpha}g_{\mu\nu} - g_{\mu\lambda}g_{\alpha\nu} )]$}}


\put(100,-10){\usebox{\gluon}}
\put(100,-10){\usebox{\rgluonarrow}}
\put(120,-10){\usebox{\gluon}}
\put(140,-10){\usebox{\gluon}}
\put(160,-10){\usebox{\gluon}}
\put(180,-10){\usebox{\lgluonarrow}}
\multiput(140,-50.5)(0,18){5}{\line(0,1){9}}
\multiput(140,-27)(-0.1,-0.1){28}{\circle*{0.2}}
\multiput(140,-27)( 0.1,-0.1){28}{\circle*{0.2}}
\multiput(140,  8)(-0.1,-0.1){28}{\circle*{0.2}}
\multiput(140,  8)( 0.1,-0.1){28}{\circle*{0.2}}
\put( 90,-10){\makebox(0,0)[r]{\small $A_{\alpha}^{a}(q)$}}
\put(150,  30){\makebox(0,0)[l]{\small $\phi_{l}(p\!+\!q\!+\!r)$}}
\put(190, -10){\makebox(0,0)[l]{\small $A_{\mu}^{m}(r)$}}
\put(150, -50){\makebox(0,0)[l]{\small $\phi_{k}(p)$}}
\put(232, -70){\makebox(0,0)[c]{(e)}}
\put(345, -10){\makebox(0,0)[c]{$-ig^{2}\{R^{a},R^{m}\}_{lk}
\,g_{\alpha\mu}$}}

\put(445,-100){\makebox(0,0)[br]{\LARGE Fig.\ 2}}

\end{picture}

\noindent
\begin{picture}(425,595)(73,-30)

\thicklines


\put(110,540){\usebox{\gluon}}
\put(110,540){\usebox{\rgluonarrow}}
\put(130,540){\usebox{\trgluon}}
\put(130,540){\usebox{\brgluon}}
\put(170,510){\line(0,1){60}}
\put(150,485){\makebox(0,0)[c]{(a)}}
\put(100,540){\makebox(0,0)[r]{\small $A_{\alpha}^{a}(q)$}}
\put(180,570){\makebox(0,0)[l]{\small $\psi_{j}(p\!+\!q)$}}
\put(180,510){\makebox(0,0)[l]{\small $\psi_{i}(p)$}}
\put(180,540){\makebox(0,0)[l]{\small $p\!-\!k$}}
\multiput(170,568)(-0.1,-0.1){28}{\circle*{0.2}}
\multiput(170,568)( 0.1,-0.1){28}{\circle*{0.2}}
\multiput(170,541)(-0.1,-0.1){28}{\circle*{0.2}}
\multiput(170,541)( 0.1,-0.1){28}{\circle*{0.2}}
\multiput(170,514)(-0.1,-0.1){28}{\circle*{0.2}}
\multiput(170,514)( 0.1,-0.1){28}{\circle*{0.2}}

\put(285,555){\makebox(0,0)[c]{pinch}}
\put(270,540){\vector(1, 0){30}}

\put(380,540){\usebox{\gluon}}
\put(380,540){\usebox{\rgluonarrow}}
\put(420,540){\usebox{\leftloop}}
\put(420,540){\usebox{\rightloop}}
\put(440,510){\line(0,1){60}}
\put(370,540){\makebox(0,0)[r]{\small $A_{\alpha}^{a}(q)$}}
\put(450,570){\makebox(0,0)[l]{\small $\psi_{j}(p\!+\!q)$}}
\put(450,510){\makebox(0,0)[l]{\small $\psi_{i}(p)$}}
\put(420,485){\makebox(0,0)[c]{(b)}}
\multiput(440,559)(-0.1,-0.1){28}{\circle*{0.2}}
\multiput(440,559)( 0.1,-0.1){28}{\circle*{0.2}}
\multiput(440,523)(-0.1,-0.1){28}{\circle*{0.2}}
\multiput(440,523)( 0.1,-0.1){28}{\circle*{0.2}}


\put(110,420){\usebox{\gluon}}
\put(110,420){\usebox{\rgluonarrow}}
\put(130,420){\usebox{\trgluon}}
\put(130,420){\usebox{\brgluon}}
\multiput(170,390.1)(0,9.2){7}{\line(0,1){4.6}}
\put(100,420){\makebox(0,0)[r]{\small $A_{\alpha}^{a}(q)$}}
\put(180,450){\makebox(0,0)[l]{\small $\phi_{l}(p\!+\!q)$}}
\put(180,390){\makebox(0,0)[l]{\small $\phi_{k}(p)$}}
\put(180,420){\makebox(0,0)[l]{\small $p\!-\!k$}}
\put(150,365){\makebox(0,0)[c]{(c)}}
\multiput(170,448)(-0.1,-0.1){28}{\circle*{0.2}}
\multiput(170,448)( 0.1,-0.1){28}{\circle*{0.2}}
\multiput(170,421)(-0.1,-0.1){28}{\circle*{0.2}}
\multiput(170,421)( 0.1,-0.1){28}{\circle*{0.2}}
\multiput(170,394)(-0.1,-0.1){28}{\circle*{0.2}}
\multiput(170,394)( 0.1,-0.1){28}{\circle*{0.2}}

\put(285,435){\makebox(0,0)[c]{pinch}}
\put(270,420){\vector(1, 0){30}}

\put(380,420){\usebox{\gluon}}
\put(380,420){\usebox{\rgluonarrow}}
\put(420,420){\usebox{\leftloop}}
\put(420,420){\usebox{\rightloop}}
\multiput(440,390.1)(0,9.2){7}{\line(0,1){4.6}}
\put(370,420){\makebox(0,0)[r]{\small $A_{\alpha}^{a}(q)$}}
\put(450,450){\makebox(0,0)[l]{\small $\phi_{l}(p\!+\!q)$}}
\put(450,390){\makebox(0,0)[l]{\small $\phi_{k}(p)$}}
\put(420,365){\makebox(0,0)[c]{(d)}}
\multiput(440,439)(-0.1,-0.1){28}{\circle*{0.2}}
\multiput(440,439)( 0.1,-0.1){28}{\circle*{0.2}}
\multiput(440,403)(-0.1,-0.1){28}{\circle*{0.2}}
\multiput(440,403)( 0.1,-0.1){28}{\circle*{0.2}}


\put(110,300){\usebox{\gluon}}
\put(110,300){\usebox{\rgluonarrow}}
\put(130,300){\usebox{\trgluon}}
\put(130,300){\usebox{\brgluon}}
\put(170,270){\usebox{\vgluon}}
\put(170,270){\usebox{\urgluonarrow}}
\put(170,290){\usebox{\vgluon}}
\put(170,290){\usebox{\ugluonarrow}}
\put(170,310){\usebox{\vgluon}}
\put(170,320){\usebox{\urgluonarrow}}
\put(100,300){\makebox(0,0)[r]{\small $A_{\alpha}^{a}(q)$}}
\put(180,330){\makebox(0,0)[l]{\small $A_{\nu}^{n}(p\!+\!q)$}}
\put(180,270){\makebox(0,0)[l]{\small $A_{\mu}^{m}(p)$}}
\put(180,300){\makebox(0,0)[l]{\small $p\!-\!k$}}
\put(150,245){\makebox(0,0)[c]{(e)}}

\put(285,315){\makebox(0,0)[c]{pinch}}
\put(270,300){\vector(1, 0){30}}

\put(380,300){\usebox{\gluon}}
\put(380,300){\usebox{\rgluonarrow}}
\put(420,300){\usebox{\leftloop}}
\put(420,300){\usebox{\rightloop}}
\put(442,270){\usebox{\vgluon}}
\put(442,270){\usebox{\ugluonarrow}}
\put(442,290){\usebox{\vgluon}}
\put(442,310){\usebox{\vgluon}}
\put(442,310){\usebox{\ugluonarrow}}
\put(370,300){\makebox(0,0)[r]{\small $A_{\alpha}^{a}(q)$}}
\put(450,330){\makebox(0,0)[l]{\small $A_{\nu}^{n}(p\!+\!q)$}}
\put(450,270){\makebox(0,0)[l]{\small $A_{\mu}^{m}(p)$}}
\put(420,245){\makebox(0,0)[c]{(f)}}


\put(110,180){\usebox{\gluon}}
\put(110,180){\usebox{\rgluonarrow}}
\put(130,180){\usebox{\trgluon}}
\put(130,180){\usebox{\brgluon}}
\put(170,150){\usebox{\vgluon}}
\put(170,150){\usebox{\urgluonarrow}}
\put(170,170){\usebox{\vgluon}}
\put(170,170){\usebox{\ugluonarrow}}
\put(170,190){\usebox{\vgluon}}
\put(170,200){\usebox{\urgluonarrow}}
\put(170,200){\usebox{\gluon}}
\put(190,200){\usebox{\lgluonarrow}}
\put(100,180){\makebox(0,0)[r]{\small $A_{\alpha}^{a}(q)$}}
\put(180,212){\makebox(0,0)[l]{\small $A_{\nu}^{n}(p\!+\!q\!+\!r)$}}
\put(180,150){\makebox(0,0)[l]{\small $A_{\mu}^{m}(p)$}}
\put(200,200){\makebox(0,0)[l]{\small $A_{\lambda}^{l}(r)$}}
\put(180,180){\makebox(0,0)[l]{\small $p\!-\!k$}}
\put(150,140){\makebox(0,0)[c]{(g)}}

\put(110,100){\usebox{\gluon}}
\put(110,100){\usebox{\rgluonarrow}}
\put(130,100){\usebox{\trgluon}}
\put(130,100){\usebox{\brgluon}}
\put(170, 70){\usebox{\vgluon}}
\put(170, 70){\usebox{\urgluonarrow}}
\put(170, 90){\usebox{\vgluon}}
\put(170, 90){\usebox{\ugluonarrow}}
\put(170,110){\usebox{\vgluon}}
\put(170,120){\usebox{\urgluonarrow}}
\put(170, 80){\usebox{\gluon}}
\put(190, 80){\usebox{\lgluonarrow}}
\put(100,100){\makebox(0,0)[r]{\small $A_{\alpha}^{a}(q)$}}
\put(180,130){\makebox(0,0)[l]{\small $A_{\nu}^{n}(p\!+\!q\!+\!r)$}}
\put(180, 68){\makebox(0,0)[l]{\small $A_{\mu}^{m}(p)$}}
\put(200, 80){\makebox(0,0)[l]{\small $A_{\lambda}^{l}(r)$}}
\put(180,100){\makebox(0,0)[l]{\small $p\!-\!k\!+\!r$}}
\put(150, 60){\makebox(0,0)[c]{(h)}}

\put(110, 20){\usebox{\gluon}}
\put(110, 20){\usebox{\rgluonarrow}}
\put(130, 20){\usebox{\trgluon}}
\put(130, 20){\usebox{\brgluon}}
\put(170,-10){\usebox{\vgluon}}
\put(170,-10){\usebox{\urgluonarrow}}
\put(170, 10){\usebox{\vgluon}}
\put(170, 10){\usebox{\ugluonarrow}}
\put(170, 30){\usebox{\vgluon}}
\put(170, 40){\usebox{\urgluonarrow}}
\put(170, 40){\usebox{\gluon}}
\put(190, 40){\usebox{\lgluonarrow}}
\put(100, 20){\makebox(0,0)[r]{\small $A_{\alpha}^{a}(q)$}}
\put(180, 52){\makebox(0,0)[l]{\small $A_{\nu}^{n}(p\!+\!q\!+\!r)$}}
\put(200, 40){\makebox(0,0)[l]{\small $A_{\mu}^{m}(p)$}}
\put(180,-10){\makebox(0,0)[l]{\small $A_{\lambda}^{l}(r)$}}
\put(180, 20){\makebox(0,0)[l]{\small $r\!-\!k$}}
\put(150,-20){\makebox(0,0)[c]{(i)}}

\put(255,100){\makebox(0,0)[c]
{$ \left. \begin{array}{c} \\ \\ \\ \\ \\ \\
\\ \\ \\ \\ \\ \\ \\ \\ \\ \end{array}\right\}$}}
\put(285,115){\makebox(0,0)[c]{pinch}}
\put(270,100){\vector(1, 0){30}}

\put(380,100){\usebox{\gluon}}
\put(380,100){\usebox{\rgluonarrow}}
\put(420,100){\usebox{\leftloop}}
\put(420,100){\usebox{\rightloop}}
\put(442, 70){\usebox{\vgluon}}
\put(442, 70){\usebox{\ugluonarrow}}
\put(442, 90){\usebox{\vgluon}}
\put(442,110){\usebox{\vgluon}}
\put(442,110){\usebox{\ugluonarrow}}
\put(440,100){\usebox{\gluon}}
\put(460,100){\usebox{\lgluonarrow}}
\put(370,100){\makebox(0,0)[r]{\small $A_{\alpha}^{a}(q)$}}
\put(450,130){\makebox(0,0)[l]{\small $A_{\nu}^{n}(p\!+\!q\!+\!r)$}}
\put(470,100){\makebox(0,0)[l]{\small $A_{\mu}^{m}(p)$}}
\put(450, 70){\makebox(0,0)[l]{\small $A_{\lambda}^{l}(r)$}}
\put(420, 45){\makebox(0,0)[c]{(j)}}

\put(500, -50){\makebox(0,0)[br]{\LARGE Fig.\ 3}}

\end{picture}

\noindent
\begin{picture}(425,260)(73,-30)

\thicklines


\put(110,180){\usebox{\gluon}}
\put(110,180){\usebox{\rgluonarrow}}
\put(130,180){\usebox{\trgluon}}
\put(130,180){\usebox{\brgluon}}
\multiput(170,150.1)(0,9.2){7}{\line(0,1){4.6}}
\put(170,200){\usebox{\gluon}}
\put(190,200){\usebox{\lgluonarrow}}
\put(100,180){\makebox(0,0)[r]{\small $A_{\alpha}^{a}(q)$}}
\put(180,212){\makebox(0,0)[l]{\small $\phi_{l}(p\!+\!q\!+\!r)$}}
\put(180,150){\makebox(0,0)[l]{\small $\phi_{k}(p)$}}
\put(200,200){\makebox(0,0)[l]{\small $A_{\mu}^{m}(r)$}}
\put(180,180){\makebox(0,0)[l]{\small $p\!-\!k$}}
\multiput(170,208)(-0.1,-0.1){28}{\circle*{0.2}}
\multiput(170,208)( 0.1,-0.1){28}{\circle*{0.2}}
\multiput(170,181)(-0.1,-0.1){28}{\circle*{0.2}}
\multiput(170,181)( 0.1,-0.1){28}{\circle*{0.2}}
\multiput(170,154)(-0.1,-0.1){28}{\circle*{0.2}}
\multiput(170,154)( 0.1,-0.1){28}{\circle*{0.2}}
\put(150,140){\makebox(0,0)[c]{(k)}}

\put(110,100){\usebox{\gluon}}
\put(110,100){\usebox{\rgluonarrow}}
\put(130,100){\usebox{\trgluon}}
\put(130,100){\usebox{\brgluon}}
\multiput(170, 70.1)(0,9.2){7}{\line(0,1){4.6}}
\put(170, 80){\usebox{\gluon}}
\put(190, 80){\usebox{\lgluonarrow}}
\put(100,100){\makebox(0,0)[r]{\small $A_{\alpha}^{a}(q)$}}
\put(180,130){\makebox(0,0)[l]{\small $\phi_{l}(p\!+\!q\!+\!r)$}}
\put(180, 68){\makebox(0,0)[l]{\small $\phi_{k}(p)$}}
\put(200, 80){\makebox(0,0)[l]{\small $A_{\mu}^{m}(r)$}}
\put(180,100){\makebox(0,0)[l]{\small $p\!-\!k\!+\!r$}}
\multiput(170,128)(-0.1,-0.1){28}{\circle*{0.2}}
\multiput(170,128)( 0.1,-0.1){28}{\circle*{0.2}}
\multiput(170,101)(-0.1,-0.1){28}{\circle*{0.2}}
\multiput(170,101)( 0.1,-0.1){28}{\circle*{0.2}}
\multiput(170, 74)(-0.1,-0.1){28}{\circle*{0.2}}
\multiput(170, 74)( 0.1,-0.1){28}{\circle*{0.2}}
\put(150, 60){\makebox(0,0)[c]{(l)}}

\put(110, 20){\usebox{\gluon}}
\put(110, 20){\usebox{\rgluonarrow}}
\put(130, 20){\usebox{\trgluon}}
\put(130, 20){\usebox{\brgluon}}
\put(170,-10){\usebox{\vgluon}}
\put(170,-10){\usebox{\urgluonarrow}}
\put(170, 10){\usebox{\vgluon}}
\put(170, 10){\usebox{\ugluonarrow}}
\put(170, 20){\usebox{\vgluon}}
\put(170, 45){\line(0,1){5}}
\put(175, 40){\line(1,0){5}}
\put(185, 40){\line(1,0){5}}
\put(100, 20){\makebox(0,0)[r]{\small $A_{\alpha}^{a}(q)$}}
\put(180, 52){\makebox(0,0)[l]{\small $\phi_{l}(p\!+\!q\!+\!r)$}}
\put(200, 40){\makebox(0,0)[l]{\small $\phi_{k}(p)$}}
\put(180,-10){\makebox(0,0)[l]{\small $A_{\mu}^{m}(r)$}}
\put(180, 20){\makebox(0,0)[l]{\small $r\!-\!k$}}
\multiput(170,48)(-0.1,-0.1){28}{\circle*{0.2}}
\multiput(170,48)( 0.1,-0.1){28}{\circle*{0.2}}
\multiput(177,40)( 0.1, 0.1){28}{\circle*{0.2}}
\multiput(177,40)( 0.1,-0.1){28}{\circle*{0.2}}
\put(150,-20){\makebox(0,0)[c]{(m)}}

\put(255,100){\makebox(0,0)[c]
{$ \left. \begin{array}{c} \\ \\ \\ \\ \\ \\
\\ \\ \\ \\ \\ \\ \\ \\ \\ \end{array}\right\}$}}
\put(285,115){\makebox(0,0)[c]{pinch}}
\put(270,100){\vector(1, 0){30}}

\put(380,100){\usebox{\gluon}}
\put(380,100){\usebox{\rgluonarrow}}
\put(420,100){\usebox{\leftloop}}
\put(420,100){\usebox{\rightloop}}
\multiput(440,70.1)(0,9.2){7}{\line(0,1){4.6}}
\put(440,100){\usebox{\gluon}}
\put(460,100){\usebox{\lgluonarrow}}
\put(370,100){\makebox(0,0)[r]{\small $A_{\alpha}^{a}(q)$}}
\put(450,130){\makebox(0,0)[l]{\small $\phi_{l}(p\!+\!q\!+\!r)$}}
\put(470,100){\makebox(0,0)[l]{\small $A_{\mu}^{m}(r)$}}
\put(450, 70){\makebox(0,0)[l]{\small $\phi_{k}(p)$}}
\multiput(440,119)(-0.1,-0.1){28}{\circle*{0.2}}
\multiput(440,119)( 0.1,-0.1){28}{\circle*{0.2}}
\multiput(440, 83)(-0.1,-0.1){28}{\circle*{0.2}}
\multiput(440, 83)( 0.1,-0.1){28}{\circle*{0.2}}
\put(420, 45){\makebox(0,0)[c]{(n)}}

\put(500, -50){\makebox(0,0)[br]{\LARGE Fig.\ 3 cont'd}}

\end{picture}

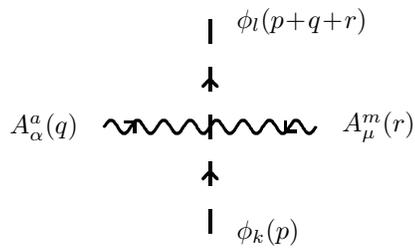
\end{document}